\documentclass{JHEP}

\usepackage{epsfig}

\makeatletter\@addtoreset{equation}{section}\makeatother

\makeatletter

\@addtoreset{equation}{section}
\makeatother

\newcommand{\be}{\begin{equation}}

\newcommand{\ee}{\end{equation}}

\newcommand{\bea}{\begin{eqnarray}}

\newcommand{\eea}{\end{eqnarray}}

\newcommand{\xx}{\nonumber\\}

\newcommand{\ct}{\cite}

\newcommand{\la}{\label}

\newcommand{\eq}[1]{(\ref{#1})}

\def\CN{{\cal N}}

\def\IR{{\hbox{{\rm I}\kern-.2em\hbox{\rm R}}}}
\def\IB{{\hbox{{\rm I}\kern-.2em\hbox{\rm B}}}}
\def\IN{{\hbox{{\rm I}\kern-.2em\hbox{\rm N}}}}
\def\IC{\,\,{\hbox{{\rm I}\kern-.59em\hbox{\bf C}}}}
\def\IZ{{\hbox{{\rm Z}\kern-.4em\hbox{\rm Z}}}}
\def\IP{{\hbox{{\rm I}\kern-.2em\hbox{\rm P}}}}
\def\IH{{\hbox{{\rm I}\kern-.4em\hbox{\rm H}}}}
\def\ID{{\hbox{{\rm I}\kern-.2em\hbox{\rm D}}}}

\def\a{\alpha}

\def\k{\kappa}

\def\rh{\rho}

\def\half{\frac{1}{2}}

\def\p{\partial}

\def\Tr{{\rm Tr}\,}

\def\ad{\dot{a}}

\def\rd{\dot{r}}
\def\sd{\dot{s}}

\def\rdp{\dot{r}^\prime}
\def\sdp{\dot{s}^\prime}

\def\rh{\hat{r}}
\def\sh{\hat{s}}

\def\rhp{\hat{r}^\prime}
\def\shp{\hat{s}^\prime}

\def\id{\dot{i}}
\def\jd{\dot{j}}

\def\idp{\dot{i}^\prime}
\def\jdp{\dot{j}^\prime}

\def\ih{\hat{i}}
\def\jh{\hat{j}}

\def\ihp{\hat{i}^\prime}
\def\jhp{\hat{j}^\prime}

\def\so{SO(4)}
\def\sop{SO(4)^\prime}

\def\bfz{{\bf Z}}

\preprint{SOGANG-HEP 311/03 \\
SNUTP 03-023 \\
{\tt hep-th/0310177}}

\author{Kyung-Seok Cha$^{\, a \,}$\footnote{quantum21@korea.com},
Bum-Hoon Lee$^{\, a \,}$\footnote{bhl@ccs.sogang.ac.kr}, and
Hyun Seok Yang$^{\, b \,}$\footnote{hsyang@phya.snu.ac.kr}
\\
${}^a${\sl Department of Physics, Sogang University,
Seoul 121-742, Korea} \\
${}^b${\sl School of Physics, Seoul National University,
Seoul 151-747, Korea}}

\title{Supersymmetric D-branes in Type IIB Plane Wave Background}

\abstract{
We systematically analyze supersymmetric D-branes in the type IIB
plane wave background using Green-Schwarz superstring theory. We
find several new supersymmetric oblique and curved D-branes. The
supersymmetries preserved by various configurations of these
D-branes including their intersections are also identified. In
particular, we show that $D_+$-branes of type $(+,-,n,n)$ for
$n=1,2,3,4$ preserve 4 dynamical supersymmetries by introducing
gauge field excitations and newly discovered oblique $D5$- and
$D7$-branes also preserve four or two dynamical supersymmetries
with gauge field condensates. Furthermore we find new curved
D-branes preserving four dynamical supersymmetries.}


\begin{document}

\renewcommand{\thefootnote}{\arabic{footnote}}
\setcounter{footnote}{0}

\section{Introduction}

In many recent developments in string theory, D-branes carrying
Ramond-Ramond charges have played an important role in the
understanding of string dualities \ct{polchinski,johnson-book}.
D-branes have precisely the correct properties to fill out duality
multiplets together with fundamental string states and other field
theoretic solitons. Furthermore D-branes have been successfully
used to explain various supersymmetric and nonsupersymmetric field
theories including the AdS/CFT correspondence and the entropy of
some black-holes. More recently, there have been attempts in
understanding some cosmological issues and the hierarchy problem
with D-branes.

D-branes can be described by boundary states of closed string
states \ct{li,gg}. The symmetries that the boundary state preserves are thus
generically the combinations of the closed string symmetries that
leave the boundary state invariant. However, it is a challenging
problem to completely classify the D-branes in a general string
background since it is necessary to quantize the string theory in
the background. Recently a maximally supersymmetric type IIB
string background was found, which is the Penrose limit of the
$AdS_5 \times S^5$ background in type IIB supergravity \ct{blau},
\bea \la{pp-metric}
&& ds^2 = -2 dx^+ dx^- - \mu^2 x_I^2 (dx^+)^2 + {dx_I}^2, \\
&& F_{+1234}=F_{+5678}=2 \mu. \nonumber
\eea
Since the string theory is exactly solvable in the Ramond-Ramond
background \eq{pp-metric} \ct{metsaev1,metsaev2},
it may be possible to get the complete spectrum of D-branes
in the background. Moreover it was realized in \ct{bmn}
that the type IIB string theory in the plane wave background
\eq{pp-metric} has a very simple description in terms of the dual
supersymmetric Yang-Mills theory.

As the first step towards this goal, we systematically classify
{\it static} D-branes in the maximally supersymmetric
type IIB plane wave background \eq{pp-metric}
using the Green-Schwarz superstring theory.
Recently it was noticed in \ct{hikida,ggsns} that
the background \eq{pp-metric} admits oblique D-branes (OD-branes)
whose isometry is a subgroup of the diagonal $SO(4)$ symmetry of
the background as well as curved D-branes preserving some
supersymmetries. The supergravity solution of oblique D-branes was
also discussed in \ct{zamaklar}.
These branes do not belong to the class of the
$D_\pm$-brane \ct{billo}-\ct{kim}.
Only a special class of oblique and curved D-branes, however,
was found in \ct{hikida,ggsns}
and it is thus demanded to know the complete list of the D-branes
in the background \eq{pp-metric} and the supersymmetries
preserved by various configurations of these D-branes.

In this paper we are using the light-cone open string theory,
where the light-cone worldvolume coordinates $X^\pm$ necessarily satisfy
the Neumann boundary condition - longitudinal branes. The
instantonic branes and branes with only one light-cone coordinate
along the worldvolume in \ct{sken-tayl} are not visible in the light-cone open
string theory. Thus these D-branes lie outside our classification.
To study these branes, one may use boundary state formalism for
the former branes and covariant gauge formulation for the latter.

This paper is organized as follows. In Sec. 2, we present a
worldsheet formulation using the Green-Schwarz superstring action
in light-cone gauge for an open string attached on a flat
$D$-brane. We first find the most general condition satisfied by
the longitudinal D-brane in the plane wave background \eq{pp-metric}
and determine the complete spectrum of the flat D-branes.
Supersymmetric flat longitudinal D-branes are summarized in Table
1.

In section 3, we give the mode expansion of open strings
consistent with a general class of open string boundary conditions
in the Green-Schwarz superstring theory context.

In Sec. 4, the analysis is generalized to the case of
intersecting D-branes using the formalism of our previous work \ct{cha}.
It is shown that oblique D-branes consistently intersect
with usual $D_\pm$-branes.

In Sec. 5, the supersymmetries preserved by various flat D-branes
are explicitly identified by finding conserved
worldsheet supercurrents consistent with open string boundary conditions \ct{kim}.
In particular, we show that $D_+$-branes of type $(+,-,n,n)$ for
$n=1,2,3,4$ preserve 4 dynamical supersymmetries by introducing
gauge field excitations and newly discovered oblique $D5$- and
$D7$-branes also preserve four or two dynamical supersymmetries
with gauge field condensates. Furthermore, we show that D-branes
with odd number of oblique directions preserve no supersymmetry
and the D-branes with even number of oblique directions preserve 8
kinematical supersymmetries of which 4 supersymmetries are
descending from the closed string and another 4 supersymmetries
are the new kind of supersymmetry, not descending from the closed
string, as identified by Skenderis and Taylor \ct{skenderis1} for $D_+$-branes.
Unbroken supersymmetry of these D-branes is summarized in Table 2.

In Sec. 6, the supersymmetry analysis is generalized to
intersecting D-branes. All supersymmetric intersecting branes are
classified in Table 3.

In Sec. 7, we also discuss supersymmetric curved D-branes
including those not mentioned in the previous literatures. We find
newly discovered curved D-branes preserve four dynamical
supersymmetries. Curved D-branes preserving dynamical
supersymmetries are listed in Table 4.

In Sec. 8, we briefly review our results obtained and discuss some
related issues.

In Appendix A, we show that Born-Infeld fluxes introduced on the
worldsheet to enhance the dynamical supersymmetry are consistent
with the equation of motion for a worldvolume gauge field on a
D5-brane.

\section{Flat D-branes in A Plane Wave Background}

The Green-Schwarz light-cone action in the plane wave background \eq{pp-metric}
describes eight free massive bosons and fermions \ct{metsaev1}. In the
light-cone gauge, $X^+=\tau$, the action is given by
\be \la{gs-action}
S = \frac{1}{2\pi \a^\prime p^+} \int d\tau \int_{0}^{2\pi \a^\prime |p^+|}
d\sigma \Bigl[ \half \p_+X_I \p_-X_I - \half \mu^2 X_I^2
- i \bar{S}(\rho^A \p_A - \mu \Pi)S \Bigr]
\ee
where $\p_\pm = \p_\tau \pm \p_\sigma$.
The equations of motion following from the action \eq{gs-action}
take the form
\bea \la{eom-boson}
&& \p_+\p_-X^I + \mu^2 X^I =0, \\
\la{eom-fermion}
&& \p_+ S^1 - \mu \Pi S^2 =0, \qquad \p_-S^2 + \mu \Pi S^1 =0.
\eea

We use the following form for $SO(8)$ gamma matrices
\be \la{8d-gamma}
\gamma^I = \pmatrix{ 0 & \gamma^I_{a\ad} \cr
                       \tilde{\gamma}^I_{\ad a} & 0 }
\ee
where $\tilde{\gamma}^I_{\ad a}=({\gamma^I}^T)_{\ad a}$
and take the $SO(8)$ chirality matrix as
\be \la{gamma9}
\gamma=\pmatrix{ 1_{8} & 0 \cr
                       0 & -1_{8} }.
\ee
In what follows, we assume that the spinors $S^A(\tau,\sigma), \; A=1,2,$
are positive chiral
fermions, $\gamma S^A = S^A$, of the form
\be \la{so8-spinor}
S_\a^A =\left(%
\begin{array}{c}
  S_a^A \\
  0 \\
\end{array}%
\right),
\ee
where $\a = 1, \cdots, 16$ and $a=1, \cdots, 8$.

Consider an open string attached on a $Dp$-brane
in the plane wave background \eq{pp-metric}.
The open string action is just defined by the action \eq{gs-action}
with string length $\a = 2\a^\prime p^+$ imposed with
appropriate boundary conditions on each end
of the open string.\footnote{\la{notation}In this paper we will
use the notation and the convention in \ct{kim} with more refined
indices. Neumann coordinates $X^r$ are decomposed into oblique directions
$X^{\rh}$ and usual parallel directions $X^{\rd}: \; r=(\rh,\rd)$.
Similarly, Dirichlet coordinates $X^{r^\prime}$ are also decomposed into
oblique directions $X^{\rhp}$ and usual parallel directions $X^{\rdp}: \;
r^\prime=(\rhp,\rdp)$.}
For longitudinal coordinates $X^r$ on D-branes without any
worldvolume flux, we impose the Neumann boundary condition
\be \la{nbc-boson}
\p_\sigma X^r|_{\p\Sigma}=0,
\ee
while for transverse coordinates $X^{r^\prime}$ we have the Dirichlet boundary
condition
\be \la{dbc-boson}
\p_\tau X^{r^\prime}|_{\p\Sigma}=0.
\ee
In the case to include gauge field excitations considered later,
some Neumann boundary conditions have to be modified as follow
\ct{sken-tayl,yoji,hikida,gaberdiel,kim}
\be \la{mnbc-boson}
(\p_\sigma X^r \pm \mu X^r)|_{\p\Sigma} =0
\ee
for some $r \in N$. The fermionic coordinates also have to satisfy the
following boundary condition at each end of the open string
\ct{lambert}
\be \la{bc-fermion}
(S^1-\Omega S^2)|_{\p\Sigma}=0,
\ee
where the matrix $\Omega$ is the products of
$\gamma$-matrices along worldvolume directions.

The boundary condition \eq{bc-fermion} has to be compatible with
the fermionic equation of motion \eq{eom-fermion} and thus
the possible type of D-branes shall be characterized by the matrix $\Gamma$
defined by
\be \la{Gamma}
\Gamma \equiv \Pi\Omega\Pi\Omega.
\ee
$D_\pm$-branes \ct{billo}-\ct{kim} are a specific class
satisfying $\Gamma = \pm 1$. Since $\Omega$ is, in general, a basis in the
Majorana representation of $SO(8)$ Clifford algebra,\footnote{\la{z2-symmetry}
The isometry in the plane wave background \eq{pp-metric} is indeed
$\so \times \sop \times {\bf Z}_2 $ where the ${\bf Z}_2 $ symmetry interchanges simultaneously
the two $SO(4)$ directions \ct{chu1}
\be \la{z2}
{\bf Z}_2  : (x^1,x^2,x^3,x^4) \leftrightarrow (x^5,x^6,x^7,x^8).
\ee
The open string theory on a D-brane is just defined by the closed string action
\eq{gs-action} by imposing the boundary conditions, \eq{nbc-boson}, \eq{dbc-boson},
and \eq{bc-fermion}. Thus the open string theory on a D-brane has to respect
the symmetry of the closed string action which is $\so \times \sop \times {\bf Z}_2 $.
Then, we think, the gluing matrix $\Omega$ defining the boundary condition
of open string fermions should be in the representation of $\so \times \sop \times {\bf Z}_2
\subset SO(8)$ Clifford algebra rather than $SO(8)$. This may explain
why OD-branes are always at $45^o$ angle in oblique directions
and the spectrum of D-branes is symmetric under the ${\bf Z}_2 $ involution \eq{z2}.} say,
$\Omega^2 = \pm 1$ and $\Pi^2 = 1$, one can see that the matrix
$\Gamma$ satisfies the following properties:
\bea \la{Gamma=Gamma}
&& \Pi \Omega \Pi \Omega = \Gamma = \Pi \Omega^T \Pi \Omega^T, \\
\la{Gamma-prop}
&& \Gamma \Gamma^T =1, \qquad \Pi \Gamma \Pi \Gamma = 1.
\eea

Since the matrix $\Gamma$ is also an element of $SO(8)$ Clifford
algebra, it must be either a symmetric or an antisymmetric matrix.
In the case the matrix $\Gamma$ is symmetric, i.e. $\Gamma^T = \Gamma$,
it follows from \eq{Gamma=Gamma} and \eq{Gamma-prop} that
\be \la{symm-Gamma}
\Gamma^2=1, \qquad [\Pi, \Gamma]=0=[\Omega, \Gamma].
\ee
The first equation in Eq. \eq{symm-Gamma} implies that $\Gamma$ is
a product of 0, 4 or 8 gamma matrices. Therefore we have three
classes of D-brane in this case:
\begin{eqnarray}\label{0-Gamma}
&& \mbox{$D_\pm$-brane}: \Gamma = \pm 1, \\
\label{2-Gamma}
&& \mbox{$ODp$-brane}: \Gamma = \pm \gamma^{i_1 i_2 i^\prime_3
i^\prime_4}, \quad (p=3,5,7), \\
\label{4-Gamma}
&& \mbox{$OD5$-brane}: \Gamma = \pm \gamma.
\end{eqnarray}
The second equation in Eq. \eq{symm-Gamma} requires that
the matrix $\Gamma$ should contain an even number of gamma matrices
in $\{ \gamma^i, \; i=1, \cdots, 4 \}$ and $\{ \gamma^{i^\prime},
\; i^\prime=5, \cdots, 8 \}$. For example,
$\Gamma = \pm \gamma^{1256}$ for $ODp$-branes in Eq. \eq{2-Gamma}.
Then one can easily find the corresponding matrix $\Omega$ for
the D-branes in Eqs. \eq{0-Gamma}-\eq{4-Gamma}. The solution shows that
the D-branes in Eqs. \eq{0-Gamma}-\eq{4-Gamma} contain 0, 2 and 4
oblique directions, respectively. Table 1 shows possible flat
D-branes with particular polarizations. Other flat D-branes
with different polarizations can be generated by $\so \times \sop$
rotations of the D-branes in Table 1. The D-branes discussed in \ct{hikida,ggsns}
correspond to the $OD3$-brane with $\Gamma = - \gamma^{1256}$
and $OD_{-}5$-brane in Table 1.

\begin{table}[tbp]
\begin{center}
\begin{tabular}{|c|c|c|c|c|} \hline
D-brane type & $\Gamma$         & $\Omega$ \\
\hline
$D_{\pm}$    & $\pm1$           & $\Omega_{D_\pm}$ \\
\hline
$OD3$        & $\pm \gamma^{1256}$ & $\frac{1}{2}(\gamma^{1}\!-\!\gamma^{6})
(\gamma^{2}\! \pm \!\gamma^{5})$ \\
\hline
             &                  & $\frac{1}{2}(\gamma^{1}\!-\!\gamma^{6})
             (\gamma^{2}\! \mp \!\gamma^{5})\gamma^{34}$ \\
$OD5$        & $\pm \gamma^{1256}$ & $\frac{1}{2}(\gamma^{1}\!-\!\gamma^{6})
(\gamma^{2}\! \mp \!\gamma^{5}) \gamma^{78}$ \\
             &                  & $\frac{1}{2}(\gamma^{1}\!-\!\gamma^{6})
             (\gamma^{2}\! \pm \!\gamma^{5})\gamma^{37}$ \\
\hline
$OD7$        & $\pm \gamma^{1256}$ & $\frac{1}{2}(\gamma^{1}\!-\!\gamma^{6})
             (\gamma^{2}\! \pm \!\gamma^{5})\gamma^{3478}$ \\
\hline
$OD_{\pm}5$  & $\pm\gamma$    & $\frac{1}{4}(\gamma^{1}\!-\!\gamma^{6})
(\gamma^{2}\! \pm \!\gamma^{5}) (\gamma^{3}\!-\!\gamma^{8})(\gamma^{4}\!+\!\gamma^{7})$ \\
\hline
\end{tabular}
\end{center}
\caption{Flat D-branes with $\Gamma^2 = 1$}
\label{tableone}
\end{table}

Applying the same argument in \ct{ggsns}, one can show that the
$OD_\pm 5$-branes in Table 1 are obtained from the usual $D_\pm 5$-branes,
respectively, by a special rotation $R$ in the coset $SO(8)/(\so \times \sop)$
which leaves the matrix $\Gamma$ invariant in the spinor space
with positive chirality. Specifically, $R$ describes a rotation by
$\pi/4$ in each of the four planes $x^1-x^6, x^2-x^5, x^3-x^8,
x^4-x^7$.

On the other hand, in the case the matrix $\Gamma$ is antisymmetric, i.e.
$\Gamma^T = - \Gamma$, it follows from \eq{Gamma=Gamma} and \eq{Gamma-prop} that
\be \la{antisymm-Gamma}
\Gamma^2= - 1, \qquad \{\Pi, \Gamma\}=0=\{\Omega, \Gamma\}.
\ee
The first equation in Eq. \eq{antisymm-Gamma} implies that $\Gamma$ is
a product of 2 or 6 gamma matrices. Thus we have two classes of
oblique D-brane in this case:
\begin{eqnarray}\label{1-Gamma}
&& \mbox{$ODp$-brane}: \Gamma = \pm \gamma^{i_1 i^\prime_2}, \quad (p=3,5,7), \\
\label{3-Gamma}
&& \mbox{$OD5$-brane}: \Gamma = \pm \gamma^{i_1 i_2 i_3 i^\prime_4
i^\prime_5 i^\prime_6}.
\end{eqnarray}
The second equation in Eq. \eq{antisymm-Gamma} forces
the matrix $\Gamma$ to contain an odd number of gamma matrices
in $\{ \gamma^i, \; i=1, \cdots, 4 \}$ and $\{ \gamma^{i^\prime},
\; i^\prime=5, \cdots, 8 \}$, for example,
$\Gamma = \pm \gamma^{16}$ for $ODp$-branes in Eq. \eq{1-Gamma}.
The matrix $\Omega$ for the D-branes in Eqs. \eq{1-Gamma}-\eq{3-Gamma}
then contain 1 and 3 oblique directions, respectively.
A few solutions of $\Omega$ are, in this case, given by
\begin{eqnarray}\label{Omega-1}
    && \Omega = \frac{1}{\sqrt{2}}(\gamma^1 \mp \gamma^6)
    \gamma^2,    \;\;\; (\Gamma = \pm \gamma^{16}), \\
    \label{Omega-3}
    && \Omega = \frac{1}{2\sqrt{2}}
    (\gamma^1 \mp \gamma^6)(\gamma^2 + \gamma^5)(\gamma^3 - \gamma^8)
    \gamma^4,    \;\;\; (\Gamma = \pm \gamma^{123568}).
\end{eqnarray}

\section{Open String Mode Expansion for D-branes}

According to the gluing matrix $\Omega$ in Table 1 and in Eqs.
\eq{Omega-1}-\eq{Omega-3}, we define diagonal coordinates
\begin{equation}\label{dia-coordinate}
    X^{\rh} = \frac{1}{\sqrt{2}}(X^r \pm X^{r^\prime}),
    \qquad X^{\rhp} = \frac{1}{\sqrt{2}}(X^{r^\prime} \mp X^r)
\end{equation}
with the index notation explained in footnote \ref{notation}.
For an OD5-brane described by $\Omega = \frac{1}{2}(\gamma^{1}\!-\!\gamma^{6})
(\gamma^{2}\! - \!\gamma^{5})\gamma^{34}$, for example, we have
\bea \la{od5-n}
&& \mbox{Neumann}: X^{\hat{1}} = \frac{1}{\sqrt{2}}(X^1 - X^{6}), \;
X^{\hat{2}} = \frac{1}{\sqrt{2}}(X^2 - X^{5}), \;
X^{\dot{3}} = X^3, \; X^{\dot{4}} = X^4, \xx
&& \mbox{Dirichlet}: X^{\hat{5}^\prime} = \frac{1}{\sqrt{2}}(X^5 + X^{2}), \;
X^{\hat{6}^\prime} = \frac{1}{\sqrt{2}}(X^6 + X^{1}), \;
X^{\dot{7}^\prime} = X^7, \; X^{\dot{8}^\prime} = X^8. \nonumber
\eea

Since the bosonic equation of motion \eq{eom-boson} is invariant
under the coordinate redefinition \eq{dia-coordinate} and
insensitive to the fermionic boundary condition \eq{bc-fermion},
the mode expansion satisfying the Neumann
or the Dirichlet boundary condition, Eqs. \eq{nbc-boson}-\eq{mnbc-boson},
is exactly the same as the usual
$D_\pm$-branes. See for this, for example, Ref. \ct{kim}.

The mode expansion of the spinor field is found to be
\bea \la{mode-spinor}
&& S^1 (\tau, \sigma)= S^1_0 (\tau, \sigma)
+ \sum_{n \neq 0}c_n(\varphi_n^1(\tau,\sigma) \Omega \widetilde{S}_n
+i \rho_n \varphi_n^2(\tau,\sigma)\Pi S_n), \xx
&& S^2 (\tau, \sigma)=  S^2_0 (\tau, \sigma)
+ \sum_{n \neq 0}c_n(\varphi_n^2(\tau,\sigma) S_n
- i \rho_n \varphi_n^1(\tau,\sigma)\Pi \Omega \widetilde{S}_n),
\eea
where the basis functions $\varphi_n^{1,2}(\tau, \sigma)$ are
defined by
\be \la{basis}
\varphi_n^1(\tau, \sigma)=e^{-i(\omega_n \tau -
\frac{n}{|\a|}\sigma)}, \qquad \varphi_n^2(\tau, \sigma)=e^{-i(\omega_n \tau
+ \frac{n}{|\a|}\sigma)}
\ee
and
\be \la{omega-etc}
\omega_n= \mbox{sign}(n) \sqrt{\mu^2 + n^2/\a^2}, \quad \rho_n=
\frac{\omega_n-n/|\a|}{\mu}, \quad c_n= \frac{1}{\sqrt{1+\rho_n^2}}.
\ee
$S^A_0 (\tau, \sigma)$ in Eq. \eq{mode-spinor} are possible zero modes
to be fixed later and the modes $\widetilde{S}_n$ are determined
by requiring the boundary condition \eq{bc-fermion} with the gluing
matrix $\Omega$ satisfying the relation \eq{Gamma}:
\begin{equation}\label{mode-sn}
    (1 + \rho_n^2 \Gamma) \widetilde{S}_n = \Bigl( (1-\rho_n^2)- i
    \rho_n \Pi \Omega (1 + \Gamma^T) \Bigr) S_n.
\end{equation}

In the case of $\Gamma^T = \Gamma$,
it is useful to decompose the spinors $S^A(\tau,\sigma)$ into
eigenspinors of $\Gamma$ by defining
\be \la{dec-s}
S^A_\pm(\tau,\sigma) = P_\pm S^A(\tau,\sigma),
\ee
where
\begin{equation}\label{projection}
    P_\pm = \half(1 \pm \Gamma).
\end{equation}
It follows from Eq. \eq{symm-Gamma} that the equations of motion
for the spinors $S_\pm^A(\tau, \sigma)$, Eq. \eq{eom-fermion},
are completely separated into two independent equations of motion
\bea \la{+eom-s+s-}
&& \p_+ S_+^1 - \mu \Pi S_+^2 =0, \qquad  \p_- S_+^2 + \mu \Pi S_+^1 =0, \\
\la{-eom-s+s-}
&& \p_+ S_-^1 -\mu \Pi S_-^2 =0, \qquad \p_- S_-^2 + \mu \Pi S_-^1=0
\eea
and the boundary condition, Eq. \eq{bc-fermion}, can be separately
imposed for the spinors $S_\pm^A(\tau, \sigma)$
\bea \la{bc-s+}
&& (S^1_+ - \Omega S^2_+)|_{\p\Sigma}=0, \\
\la{bc-s-}
&& (S^1_- - \Omega S^2_-)|_{\p\Sigma}=0.
\eea
Then one can immediately see from Eqs. \eq{mode-spinor} and
\eq{mode-sn} that the spinor $S_+^A(\tau, \sigma)$ has a
$D_+$-like mode expansion while $S_-^A(\tau, \sigma)$ does a
$D_-$-like mode expansion \ct{kim}. This should be obvious since $\Gamma = +1$
in the space spanned by $S_+^A$ while $\Gamma = -1$
in the space spanned by $S_-^A$, namely,
\be \la{eigen-Gamma}
\Gamma S^A_\pm(\tau,\sigma) = \pm S^A_\pm(\tau,\sigma).
\ee

Based on this observation, we can easily find the zero modes
$S^A_0 (\tau, \sigma) =  S^A_{-0} (\tau, \sigma) + S^A_{+0} (\tau, \sigma)$
in Eq. \eq{mode-spinor}:
\begin{eqnarray}\label{zero-mode}
 S^1_0 (\tau, \sigma) &=& \cos \mu\tau S_0^- - \sin\mu\tau \Omega \Pi S_0^-
 + \cosh \mu\sigma S_0^+ + \sinh \mu\sigma \Omega \Pi S_0^+, \xx
 S^2_0 (\tau, \sigma) &=& \cos \mu\tau \Omega^T S_0^-
- \sin\mu\tau \Pi S_0^- + \cosh \mu\sigma \Omega^T S_0^+
+ \sinh \mu\sigma \Pi S_0^+,
\end{eqnarray}
where
\begin{equation}\label{s0}
    P_\pm S^\pm_0 = S^\pm_0, \qquad  P_\mp S^\pm_0 = 0.
\end{equation}

The commutation relations between the modes read as
\bea \la{com-rel}
&& \{ S_0 ^{-a}, S_0^{-b} \} = \frac{1}{4} P_-^{ab}, \xx
&& \{ S_0^{+a}, S_0^{+b} \} = \frac{\pi \mu |\a|}{4\sinh \pi \mu |\a|}
\Bigl(P_+^{ab}\cosh\pi\mu|\a|-(P_+\Omega \Pi)^{ab}\sinh\pi\mu|\a| \Bigr), \xx
&& \{S_n^{\pm a}, S_m^{\pm b} \} = \frac{1}{4} \delta_{n+m,0}
P_\pm^{ab},
\eea
where $S_n^{\pm} = P_\pm S_n$ for $n \neq 0$.

One can show that there is
no spinor zero mode in the case of $\Gamma^T = - \Gamma$,
that is, $S^A_0 (\tau, \sigma)=0$.
This fact signals no kinematical supersymmetry in this case. In
section 5 we will indeed show that the OD-branes in this case,
Eqs. \eq{1-Gamma}-\eq{3-Gamma},
preserve no supersymmetry. Thus we will not give a detailed
analysis for this class of OD-branes.

\section{Intersecting D-branes}

In this section, we will generalize the previous analysis to the
case of intersecting D-branes using the formalism in \ct{cha}.
That is, we now consider an open string stretched between
$Dp$-brane and $Dq$-brane with appropriate boundary conditions on
each end of the open string in the plane wave background \eq{pp-metric}.
In particular, the fermionic coordinates have to satisfy the
following boundary condition at each end of the open string
\be \la{ibc-fermion}
(S^1-\Omega_0 S^2)|_{\sigma=0}=0, \qquad
(S^1- \Omega_\pi S^2)|_{\sigma=\pi \a}=0,
\ee
with the matrix $\Omega_\theta =(\Omega_0, \Omega_\pi)$
satisfying
\be \la{Gamma-0pi}
\Pi\Omega_0 \Pi\Omega_0 = \Gamma_0, \qquad
\Pi\Omega_\pi \Pi \Omega_\pi = \Gamma_\pi.
\ee
Here the D-brane is either a $D_\pm$-brane or an OD-brane.

The coordinates $X^I(\tau, \sigma)$ of a $p-q$ string
can be partitioned into four sets, NN, DD, ND, and DN, according
to whether the coordinate $X^I$ has Neumann (N) or Dirichlet (D)
boundary condition at each end.\footnote{\la{inter-index}
For intersecting D-branes, we will use indices
 $(r, s, \cdots)=(\rh, \sh, \cdots; \rd, \sd, \cdots)
 \;(r^\prime, s^\prime,\cdots)= (\rhp, \shp, \cdots; \rdp, \sdp, \cdots),
\; (i,j,\cdots)=(\ih, \jh, \cdots; \id, \jd, \cdots)$, and
$(i^\prime,j^\prime, \cdots)=(\ihp, \jhp, \cdots; \idp, \jdp, \cdots)$
for NN, DD, ND, and DN coordinates, respectively, with a
distinction between hatted indices for oblique directions and
dotted indices for parallel directions.} For the same reason in
section 3, the bosonic coordinates have the same mode expansion
as the case of $D_\pm$-brane intersection.

The mode expansion of the spinor field can be determined by
exactly the same method as that in \ct{cha}.
We take an appropriate combination of spinor
fields $\xi^A(\tau, \sigma)$ with integer modes and
$\eta^A(\tau,\sigma)$ with half-integer modes
or with ${\bf R}$-modes to be compatible with supersymmetry:
\bea \la{spinor-inter}
&& S^1(\tau,\sigma)=\left\{%
\begin{array}{ll}
    I_+ \xi^1(\tau,\sigma) + I_- \eta^1(\tau,\sigma), & \;\; \hbox{for A-type;} \\
    I_- \xi^1(\tau,\sigma) + I_+ \eta^1(\tau,\sigma), & \;\; \hbox{for B-type,} \\
\end{array}%
\right. \xx
&& S^2(\tau,\sigma)=I_+ \xi^2(\tau,\sigma) + I_-
\eta^2(\tau,\sigma),
\eea
where $16 \times 16$ matrices $I_+$ and $I_-$ are defined by
\be \la{i+-}
I_+ =  \half(1 + \Omega_0^T \Omega_\pi), \qquad
I_- =  \half(1 - \Omega_0^T \Omega_\pi).
\ee
In Eq. \eq{spinor-inter}, the A-type solution is for $|p-q|=0,4,8$ in $Dp-Dq$ brane
configurations while the B-type solution for $|p-q|=2,6$.
The spinors $\xi^A(\tau,\sigma)$ and $\eta^A(\tau,\sigma)$
are taken as the solution of the equations of motion
\eq{eom-fermion} satisfying the boundary condition \eq{ibc-fermion}
at $\sigma =0$:
\bea \la{open-fermion}
&& \xi^1 (\tau, \sigma)= S^1_0(\tau, \sigma)
+ \sum_{n \neq 0}c_n(\varphi_n^1(\tau, \sigma) \Omega_0 \widetilde{S}_n
+i \rho_n \varphi_n^2(\tau, \sigma)\Pi S_n), \xx
&& \xi^2 (\tau, \sigma)= S^2_0(\tau, \sigma)
+ \sum_{n \neq 0}c_n(\varphi_n^2(\tau, \sigma) S_n
- i \rho_n \varphi_n^1(\tau, \sigma)\Pi \Omega_0 \widetilde{S}_n), \xx
&& \eta^1 (\tau, \sigma)= \sum_{\k }c_\k(\varphi_\k^1(\tau, \sigma)
\Omega_0 \widetilde{S}_\k +i \rho_\k \varphi_\k^2(\tau, \sigma)\Pi S_\k), \xx
&& \eta^2 (\tau, \sigma)= \sum_{\k }c_\k (\varphi_\k^2(\tau, \sigma)
S_\k - i \rho_\k \varphi_\k^1(\tau, \sigma)\Pi \Omega_0 \widetilde{S}_\k),
\eea
where the basis functions $\varphi_\nu^{1,2}(\tau, \sigma)$ are defined by
\be \la{i-basis}
\varphi_\nu^1(\tau, \sigma)=e^{-i(\omega_\nu \tau -
\frac{\nu}{|\a|}\sigma)}, \qquad \varphi_\nu^2(\tau, \sigma)=e^{-i(\omega_\nu \tau
+ \frac{\nu}{|\a|}\sigma)}
\ee
and
\be \la{i-omega-etc}
\omega_\nu= \mbox{sign}(\nu) \sqrt{\mu^2 + \nu^2/\a^2}, \quad \rho_\nu=
\frac{\omega_\nu-\nu/|\a|}{\mu}, \quad c_\nu= \frac{1}{\sqrt{1+\rho_\nu^2}}
\ee
for $\nu = n \in \bfz$ or $\k \in \bfz + \half, \bf{R}$.
Here $S^A_0(\tau,\sigma)$ in the case of $\Gamma^T_0 = \Gamma_0$
are given by Eq. \eq{zero-mode} with
a replacement $(\Omega, \Gamma) \to (\Omega_0, \Gamma_0)$
while $S^A_0(\tau,\sigma)=0$ in the case of $\Gamma^T_0 = - \Gamma_0$ and
\begin{equation}\label{imode-sn}
    (1 + \rho_\nu^2 \Gamma_0 ) \widetilde{S}_\nu = \Bigl( (1-\rho_\nu^2)
    - i \rho_\nu \Pi \Omega_0 (1 + \Gamma_0^T) \Bigr) S_\nu.
\end{equation}

We now require the spinors $S^A(\tau,\sigma)$ in Eq.
\eq{spinor-inter} to satisfy the equations of motion
\eq{eom-fermion} and then we need the following condition on
$I_\pm$:
\be \la{i+-pi}
\Pi I_\pm = \left\{%
\begin{array}{ll}
    I_\pm \Pi, &  \;\; \hbox{for A-type;} \\
    I_\mp \Pi, & \;\; \hbox{for B-type.} \\
\end{array}%
\right.
\ee
Noting that the matrices in Eq. \eq{i+-pi} are acting on the
positive chirality spinors $S^A(\tau,\sigma)$, one can see that
the condition \eq{i+-pi} is equivalent to the following constraint
\be \la{pi-Gamma}
\Gamma_0 \Gamma_\pi  = \left\{%
\begin{array}{ll}
    1 \; \mbox{or} \; \gamma,    &  \;\; \hbox{for} \; \Gamma^T_\theta =\Gamma_\theta ; \\
   -1 \; \mbox{or} \; -\gamma,   &  \;\; \hbox{for} \; \Gamma^T_\theta =-\Gamma_\theta. \\
\end{array}%
\right.
\ee
The condition \eq{pi-Gamma} clearly explains why a $D_-$-brane cannot
have a supersymmetric intersection with a $D_+$-brane, as was
shown in \ct{cha}, since $\Gamma_0 =-1$ and $\Gamma_\pi =1$
for this kind of intersection. In addition, the condition \eq{pi-Gamma}
implies that there may be a supersymmetric intersection between
different classes of OD-brane or an $OD_{\pm}5$-brane and
a $D_{\pm}p$-brane only if they satisfy $\Gamma_0 \Gamma_\pi =
\gamma$. In section 6 we will show that this case preserves
only kinematical supersymmetries.

Note that $\Omega^T_\theta = - \Omega_\theta$ for $D3$- and $D7$-branes,
but $\Omega^T_\theta = \Omega_\theta$ for $D5$-branes and thus
\be \la{ab-type}
\Omega_0^T \Omega_\pi = \left\{%
\begin{array}{ll}
    \Omega_0 \Omega_\pi^T, & \;\; \hbox{for A-type;} \\
    - \Omega_0 \Omega_\pi^T, & \;\; \hbox{for B-type.} \\
\end{array}%
\right.
\ee
Using Eq. \eq{ab-type}, we get useful identities \ct{cha}:
\bea \la{identity-a}
&& \Omega_\theta I_\pm = I_\pm \Omega_\theta,
\quad I_\pm \Omega_0 = \pm  I_\pm
\Omega_\pi, \qquad \mbox{for A-type;} \\
\la{identity-b}
&& \Omega_\theta I_\pm = I_\mp \Omega_\theta,
\quad I_\pm \Omega_0 = \mp  I_\pm
\Omega_\pi, \qquad \mbox{for B-type}.
\eea
It is not difficult to check using Eqs. \eq{identity-a}-\eq{identity-b}
that the spinors in Eq. \eq{spinor-inter} satisfy
the boundary conditions \eq{ibc-fermion} only if the mode number $\k$ satisfies
the following equation
\begin{equation}\label{mode-k}
    e^{2\pi i \k} \Bigl( (1- \rho_\k^2) - i \rho_\k \Pi \Omega_0
    (1 + \Gamma_0^T) \Bigr) S_\k = - \Bigl( (1- \rho_\k^2) + i \rho_\k \Pi \Omega_0
    (1 + \Gamma_0^T) \Bigr) S_\k.
\end{equation}
For example, $\k  \in \bfz + \half$ when $\Gamma_0 = -1$
and when $\Gamma_0 = 1$ \ct{cha}
\be \la{k=+1}
e^{2\pi i \k} = - \frac{\k + i \mu |\a|}{\k - i \mu |\a|} \quad
\mbox{or} \quad - \frac{\k - i \mu |\a|}{\k + i \mu |\a|}.
\ee

The matrix $\Omega_0^T \Omega_\pi$ consists of products of $\gamma$-matrices along the
ND and DN directions. Since $(\Omega_0^T \Omega_\pi)^2=1$
and $\Tr(\Omega_0^T \Omega_\pi)=0$
for $\Omega_0^T \Omega_\pi \neq \pm 1$, there can
be only three kinds of possibility:
\be \la{omega-sol}
\Omega_0^T \Omega_\pi= \left\{%
\begin{array}{ll}
    \pm 1, & \;\; \hbox{$\sharp_{ND}=0$}, \\
    \pm \gamma, & \;\; \hbox{$\sharp_{ND}=8$}, \\
    \pm \pmatrix{
  \Xi & 0 \cr
  0 & \pm \Xi \cr}, & \;\; \hbox{$\sharp_{ND}=4$}, \\
\end{array}%
\right.
\ee
where
\be \la{Xi}
\Xi = \pmatrix{
  1_4 & 0 \cr
  0 & -1_4 \cr},
\ee
and $\sharp_{ND}$ denotes the total number of ND and DN directions.

The case $\Omega_0^T \Omega_\pi=1$ corresponds to
parallel $Dp$-branes while the case $\Omega_0^T
\Omega_\pi=-1$ corresponds to $Dp$-anti-$Dp$ branes, but
the cases $\Omega_0^T \Omega_\pi=\pm \gamma$ and $\Omega_0^T
\Omega_\pi=\pm \Xi$ correspond to $Dp-Dq$ or $Dp$-anti-$Dq$
branes with $\sharp_{ND}=8$ and $\sharp_{ND}=4$, respectively.
Note that the B-type branes allow only the $\sharp_{ND}=4$ case.

\section{Supersymmetry of Flat D-branes}

In a light-cone gauge, the 32 components of the supersymmetries
for a closed string decompose into kinematical supercharges,
$Q^{+A}_a$, and dynamical supercharges, $Q^{-A}_{\ad}$.
For a closed superstring in the plane wave background with
the action \eq{gs-action}, the conserved super-N\"other
charges were identified by Metsaev \ct{metsaev1}:
\bea \la{charge-p}
&& Q^{+1} = \frac{\sqrt{2p^+}}{2\pi \a^\prime p^+}
\int_{0}^{2\pi \a^\prime |p^+|} d\sigma (\cos\mu\tau S^1 -\sin \mu\tau \Pi S^2), \\
\la{charge-q+2}
&& Q^{+2} = \frac{\sqrt{2p^+}}{2\pi \a^\prime p^+}
\int_{0}^{2\pi \a^\prime |p^+|} d\sigma (\cos\mu\tau S^2 + \sin \mu\tau \Pi S^1), \\
\la{charge-q-1}
&& \sqrt{2p^+}Q^{-1}= \frac{1}{2\pi \a^\prime p^+}
\int_{0}^{2\pi \a^\prime |p^+|} d\sigma \Bigl( \p_- X^I\gamma^I S^1
-\mu X^I\gamma^I \Pi S^2 \Bigr), \\
\la{charge-q-2}
&& \sqrt{2p^+}Q^{-2}= \frac{1}{2\pi \a^\prime p^+}
\int_{0}^{2\pi \a^\prime |p^+|} d\sigma \Bigl( \p_+ X^I\gamma^I
S^2 + \mu X^I\gamma^I \Pi S^1 \Bigr).
\eea

The kinematical supersymmetry is, in general, related to a shift
of spinor fields and thus generated by spinor zero modes. We
showed in Eq. \eq{zero-mode} that there are two kinds of spinor
zero modes, $S_0^\pm$, when $\Gamma^T = \Gamma$. Therefore we
expect that there are two kinds of kinematical supersymmetry in
this case where each of supersymmetry is generated by $S_0^\pm$.
Indeed we will show that an open string on a D-brane with even
number of oblique directions preserves 8 kinematical
supersymmetries of which 4 supersymmetries generated by $S_0^-$
are descending from the closed string and another 4
supersymmetries generated by $S_0^+$ are the new kind of
supersymmetry, not descending from the closed string, as
identified by Skenderis and Taylor \ct{skenderis1} for $D_+$-branes.

Indeed it was shown in \ct{skenderis1,skenderis2} that
the light-cone action admits an infinite number of worldsheet
symmetries and they all act on the open string spectrum. Since the
worldsheet theory on OD-branes can be completely separable
into $D_\pm$-like ones as was shown in Eqs.
\eq{+eom-s+s-}-\eq{bc-s-}, it is obvious that the worldsheet
symmetries in \ct{skenderis1,skenderis2} are also applicable to
the OD-branes.  Since we are interested in the open string
supersymmetry, we will focus only on the kinematical supersymmetry
generated by spinor zero modes $S_0^+$.

To see this, let us first consider the kinematical supersymmetry
descending from the closed string
\begin{equation}\label{ks-}
q^+_{-} = Q^{+1} + \Omega Q^{+2}
\end{equation}
from which the N\"other charge density $q^+_{-\tau}$ reads as
\begin{equation}\label{ks-density}
q^+_{-\tau} = \sqrt{2p^+} \Bigl( \cos \mu \tau (S^1 + \Omega S^2)
+ \sin \mu \tau \Omega\Pi (S^1 - \Gamma\Omega S^2) \Bigr).
\end{equation}
It is easy to show using the equations of motion, Eqs. \eq{eom-boson}
and \eq{eom-fermion}, that the kinematical supercharge density $q^+_{-\tau}$
satisfies the following conservation law
\be \la{con-law-kin-}
\frac{\p q^+_{-\tau}}{\p \tau} + \frac{\p q^+_{-\sigma}}
{\p \sigma} =0
\ee
with
\be \la{current-kin-}
q^+_{-\sigma} = \sqrt{2p^+} \Bigl( \cos \mu\tau (S^1 - \Omega S^2)
+ \sin \mu\tau \Omega\Pi (S^1 + \Gamma \Omega S^2)
\Bigr).
\ee
One can immediately see from Eqs. \eq{current-kin-} and
\eq{eigen-Gamma} that the
kinematical supersymmetry defined by
\begin{equation}\label{ks-d-}
    q^+_{D_-} = P_- q^+_{-}
\end{equation}
is strictly preserved and is generated by the zero mode $S_0^-$.

Next consider the kinematical supersymmetry, not descending from
the closed string, preserved on $D_+$-branes
\begin{equation}\label{ks+}
q^+_{+} = \frac{1}{\pi |\a|} \int^{\pi |\a|}_0 d \sigma q_{+\tau}^+
\end{equation}
where
\begin{equation}\label{ks+density}
q^+_{+\tau} = \sqrt{2p^+} \sqrt{\frac{\pi \mu |\a|}{\sinh \pi \mu |\a|}} \,
e^{\mu(\sigma-\half \pi|\a|) \Omega \Pi} (S^1 + \Omega S^2).
\ee
It is also easy to show that
\bea \la{con-law-kin+}
\frac{\p q^+_{+\tau}}{\p \tau} + \frac{\p q^+_{+\sigma}}
{\p \sigma} &=& \mu \sqrt{2p^+} \sqrt{\frac{\pi \mu |\a|}{\sinh \pi \mu |\a|}} \,
e^{\mu(\sigma-\half \pi|\a|) \Omega \Pi} \Omega\Pi \times \xx
&& \Bigl( (S^1 - \Omega S^2) - (S^1 - \Gamma \Omega S^2)
\Bigr)
\eea
where
\be \la{current-kin+}
q^+_{+\sigma} = \sqrt{2p^+} \sqrt{\frac{\pi \mu |\a|}{\sinh \pi \mu |\a|}} \,
e^{\mu(\sigma-\half \pi|\a|) \Omega \Pi} (S^1 - \Omega S^2).
\ee
It is thus obvious that the kinematical supersymmetry defined by
\begin{equation}\label{ks-d+}
    q^+_{D_+} = P_+ q^+_{+}
\end{equation}
is preserved and is generated by the zero mode $S_0^+$.

In consequence, an open string on a D-brane satisfying $\Gamma^T =
\Gamma$ preserves 8 kinematical supersymmetries of which 4 supersymmetries
are generated by $S_0^-$ and another 4 supersymmetries are
generated by $S_0^+$. We summarized our results on the
supersymmetry of D-branes in Table 2.

On the other hand, an open string on a D-brane satisfying $\Gamma^T
= - \Gamma$ preserves no kinematical supersymmetry since the
matrix $\Gamma$ has eigenvalues $\pm i$ in this case and
thus $q^+_{-\sigma}|_{\p \Sigma}$ in Eq. \eq{current-kin-}
and the right hand side of Eq. \eq{con-law-kin+} cannot vanish
for any component of spinors. This is consistent with the fact
that there is no spinor zero mode when $\Gamma^T
= - \Gamma$.

Now we investigate the dynamical supersymmetry preserved by an
open string on a D-brane characterized by $\Gamma$ in Eq. \eq{Gamma}.
The dynamical supercharge of an open string is given by a
combination of those of a closed string compatible with the open
string boundary conditions.
Due to the boundary condition \eq{bc-fermion},
it turns out that the conserved dynamical supercharge is given by
(a subset of)\footnote{\la{diff-def}Note that the current definition of the dynamical
supercharge $q^-$ differs by a factor $-\Omega$ from that in
\ct{kim,cha}.}
\be \la{dyn-susy}
q^- = Q^{-1}- \Omega Q^{-2}.
\ee
Using the similar recipe used in the kinematical
supersymmetry, it is not difficult to show that the dynamical supercharge density
$q^-_\tau$ in Eq. \eq{dyn-susy} also satisfies the conservation law
\be \la{con-law-dyn}
\frac{\p q^-_\tau}{\p \tau} + \frac{\p q^-_\sigma}
{\p \sigma} = 0,
\ee
where
\bea \la{ds-current}
q^-_\sigma &=& \sqrt{\frac{1}{2p^+}}
\Bigl( (\p_\tau X^r \gamma^r - \p_\sigma X^{r^\prime} \gamma^{r^\prime})
(S^1 - \Omega S^2) \xx
&& + (\p_\tau X^{r^\prime} \gamma^{r^\prime} - \p_\sigma X^r \gamma^r)
(S^1 + \Omega S^2) \xx
&& + \mu X^r \gamma^r \Omega \Pi (S^1 + \Gamma \Omega S^2)
- \mu X^{r^\prime} \gamma^{r^\prime} \Omega \Pi (S^1 - \Gamma
\Omega  S^2) \Bigr).
\eea

If $\Gamma = \pm 1$, we definitely recover the $D_\pm$-brane
case \ct{kim}. It was shown in \ct{sken-tayl} that $D_-$-branes of type
$(+,-,3,1)$ or $(+,-,1,3)$ with a constant worldvolume flux also
preserve 16 supersymmetries whose possibility was not discussed in
\ct{kim}. Here we will show how it can be true. In the presence of
a constant worldvolume flux $F_{+ r}$, there is an additional
boundary action given by
\begin{equation}\label{bd-action}
S_B = \frac{1}{2\pi \a^\prime p^+} \int d\tau \int_{0}^{2\pi \a^\prime |p^+|}
d\sigma F_{+ r} \p_\tau X^+ \p_\sigma X^{r} =
\frac{F_{+ r}}{2 \pi \a^\prime p^+} \int_{\p \Sigma}  d\tau X^r.
\end{equation}
This boundary term affects the Neumann boundary condition as
follows
\begin{equation}\label{d-flux-bc}
    \p_\sigma X_r - F_{+ r} = 0.
\end{equation}
With this boundary condition, only $D5$-brane among $D_-$-branes
can preserve dynamical supersymmetry. To see this, let us take a
$(+,-,3,1)$-brane, for definiteness, extended along
$(+,-,1,2,3,5)$ directions, say, $N= (1,2,3,5)$ and $ D=
(4,6,7,8)$ and thus we have $\Gamma = -1$ and $\Omega\Pi = - \gamma^{45}$
in this case. If we turn on the flux $F_{+5} = \mu q$ only, otherwise $ F_{+ r} =
0$, the current $q^-_\sigma$ in Eq. \eq{ds-current} at the
boundary reduces to
\be \la{flux-bcurrent}
q^-_\sigma|_{\p \Sigma} = - \sqrt{\frac{1}{2p^+}}
(\p_\sigma X^5 -  \mu X^4) \gamma^5 (S^1 + \Omega S^2),
\ee
where we used $\p_\sigma X^r|_{\p \Sigma} = 0, \; r=1,2,3$ and
$X^{r^\prime}|_{\p \Sigma}=0, \; r^\prime = 6,7,8$.
Thus we see that $q^-_\sigma|_{\p \Sigma}$ identically vanishes if
$X^4|_{\p \Sigma} = q$, as was shown in \ct{sken-tayl}. From the
above analysis, we see that the condition $\gamma^r =
\pm \gamma^{r^\prime} \Omega\Pi$ is necessary to preserve the
dynamical supersymmetry with a constant flux. Obviously
$(+,-,3,1)$- and $(+,-,1,3)$-branes only satisfy this condition.

When $\Gamma = \Omega \Pi\Omega \Pi=1$, there is also
a new possibility for $D_+$-branes of type $(+,-,n,n)$ with
$n=1,2,3,4$ to preserve dynamical supersymmetries by introducing a
gauge field excitation,\footnote{This idea was arisen from a discussion
with Yasuaki Hikida who we thank for the helpful discussion.}
whose possibility was anticipated by Hikida and Yamaguchi \ct{hikida}
from general supersymmetry arguments. To prove this claim, we
first introduce projection matrices defined by
\begin{equation}\label{proj-d+}
    P_\pm^{D_+} = \half ( 1 \pm \Omega \Pi)
\end{equation}
and projected supercharges
\be \la{proj-d+susy}
q^-_\pm \equiv P_\pm^{D_+} (Q^{-1}- \Omega Q^{-2}).
\ee

To proceed our argument, it is convenient to decompose Neumann and
Dirichlet coordinates into two $SO(4)$ directions:
$r=(r_1,r_2)$ and $r^\prime=(r_1^\prime, r_2^\prime)$ where the
subscripts $1 \in SO(4)$ and $2 \in SO(4)^\prime$ are used for that purpose.
From the definition \eq{proj-d+}, we get
\begin{eqnarray}\label{d+p+-comm1}
&& [P_\pm^{D_+},\gamma^{r_1}]=0=[P_\pm^{D_+},\gamma^{r^\prime_2}], \\
\label{d+p+-comm2}
&& P_\pm^{D_+} \gamma^{r_2}=\gamma^{r_2}P_\mp^{D_+}, \quad
P_\pm^{D_+} \gamma^{r_1^\prime}=\gamma^{r_1^{\prime}}P_\mp^{D_+}.
\end{eqnarray}
From Eq. \eq{ds-current} using Eqs. \eq{d+p+-comm1}-\eq{d+p+-comm2},
it is easy to find the condition for the projected supercharge
\eq{proj-d+susy} to be conserved since $q^-_{\pm \sigma}$ at the
worldsheet boundary $\p \Sigma$ reduces to
\begin{eqnarray}\label{d+q+sigma}
q^-_{\pm \sigma} \bigg|_{\partial\Sigma}&=& \sqrt{\frac{1}{2p^+}}
\Bigl( (\partial_\tau X^{r_2^\prime} \gamma^{r_2^\prime} -
\partial_\sigma X^{r_1} \gamma^{r_1}
\pm \mu X^{r_1} \gamma^{r_1}) P_\pm^{D_+}(S^1 + \Omega S^2) \xx
&& + (\partial_\tau X^{r_1^\prime} \gamma^{r_1^\prime} -
\partial_\sigma X^{r_2} \gamma^{r_2}
\mp \mu X^{r_2} \gamma^{r_2}) P_\mp^{D_+}(S^1 + \Omega S^2)
\Bigr)_{\partial\Sigma}.
\end{eqnarray}
We see that the Neumann boundary condition should be modified as
follows
\be \la{mbc-d+}
\Bigl(\partial_\sigma X^{r_1}
\mp \mu X^{r_1} \Bigr)_{\partial\Sigma} = 0 =
\Bigl(\partial_\sigma X^{r_2}
\pm \mu X^{r_2} \Bigr)_{\partial\Sigma}
\ee
to preserve the dynamical supersymmetry. This kind of boundary
condition can be realized by introducing a gauge field excitation
of the form
\be \la{d+boundary}
S_B = \frac{1}{2\pi \a^\prime p^+} \int d\tau \int_{0}^{2\pi \a^\prime |p^+|}
d\sigma F^{-r} \p_\sigma X^{r} =
\frac{\mu}{4 \pi \a^\prime p^+} \int_{\p \Sigma}  d\tau
(\pm X^{r_1} X^{r_1} \mp X^{r_2} X^{r_2}),
\ee
where the Born-Infeld flux $F^{-r}$ satisfies
\be \la{d+bi}
\p_r F^{-r}= 0, \qquad r = (r_1,r_2).
\ee

Therefore we proved that $D_+$-branes of type $(+,-,n,n)$ with
$n=1,2,3,4$ preserve 4 dynamical supersymmetries by introducing a
gauge field excitation, consistent with the result in \ct{hikida}.
Note that the dynamical supersymmetry in this case is preserved
regardless of transverse locations of D-brane.
One can check that Eq. \eq{d+bi} is just the equation of motion
for the gauge field $A$, Eq. \eq{d5eom-A} with $F_5 = 0$.
In this case we don't need an additional Chern-Simons coupling
such as \eq{Chern-Simons}.

When $\Gamma^T = - \Gamma$, however, Eq.
\eq{ds-current} immediately shows that there is no chance for
$q^-_{\sigma}|_{\p \Sigma}$ to vanish and thus an open string on
this D-brane does not preserve any dynamical supersymmetry at all.

In the case of $\Gamma^T = \Gamma$, on the other hand, the
condition for $q^-_{\sigma}|_{\p \Sigma}$ to vanish depends on the
eigenvalue of the matrix $\Gamma$. We therefore introduce
projected supercharges defined by
\be \la{projdyn-susy}
q^-_\pm \equiv P_\pm (Q^{-1}- \Omega Q^{-2}).
\ee
In particular, when $\Gamma = - \gamma$, eight $q^-_+$ supersymmetries
survive only if $X^{\rhp} =0 $ for all $\rhp \in D$
while $q^-_-$ identically vanish because
$[\gamma, \Omega]=[\gamma, \Pi]=0$ and $\gamma S^A = S^A$.
This might be expected since the $OD_\pm 5$-branes
can be related to the ordinary $D_\pm 5$-branes
by a special rotation $R$ in the coset $SO(8)/(\so \times \sop)$
which leaves the matrix $\Gamma$ invariant in the spinor space
with positive chirality \ct{ggsns}.

One may wonder whether a $D_+5$-brane with flux, preserving 8
dynamical supersymmetries, can be rotated to an $OD_+ 5$-brane
preserving the same kind of supersymmetry. In this case the
Neumann boundary condition of the $OD_+5$-brane needs to be modified
to $\p_\sigma X^r - \mu X^r =0, \; \forall r \in N$.
The general expression \eq{ds-current}, however, implies that it is not possible
since we need $\Omega\Pi S^A = S^A$ for $OD5$-branes,
which is never satisfied. It will be shown in Appendix A that this
can also be understood by investigating a worldvolume Chern-Simons
coupling.

To find the condition for the projected supercharge to be
conserved, first note that
\begin{eqnarray}\label{p+-comm1}
&& P_\pm \gamma^{\hat{r}}=\gamma^{\hat{r}}P_\mp, \quad
P_\pm \gamma^{\hat{r^\prime}}=\gamma^{\hat{r^{\prime}}}P_\mp, \\
\label{p+-comm2}
&& [P_\pm,\gamma^{\rd}]=0=[P_\pm,\gamma^{\rd^\prime}].
\end{eqnarray}
It is then easy to read off the value of $q^-_{\pm \sigma}|_{\p
\Sigma}$ from Eq. \eq{ds-current} using Eqs. \eq{p+-comm1}-\eq{p+-comm2}:
\begin{eqnarray}\label{dynq+sigma}
q^-_{+\sigma} \bigg|_{\partial\Sigma}&=& - \sqrt{\frac{1}{2p^+}}
\Bigl( (\partial_\sigma X^{\rd} \gamma^{\rd}
- \mu X^{\rd} \gamma^{\rd } \Omega \Pi) (S^1_+ + \Omega S^2_+) \xx
&& + (\partial_\sigma X^{\hat{r}}\gamma^{\hat{r}}
+ \mu X^{\rhp}\gamma^{\rhp}\Omega \Pi) (S^1_- + \Omega S^2_-)
\Bigr)_{\partial\Sigma}
\end{eqnarray}
and
\begin{eqnarray}\label{dynq-sigma}
q^-_{-\sigma} \bigg|_{\partial\Sigma}&=& - \sqrt{\frac{1}{2p^+}}
\Bigl( (\partial_\sigma X^{\rh} \gamma^{\rh}
- \mu X^{\rh} \gamma^{\rh} \Omega \Pi) (S^1_+ + \Omega S^2_+) \xx
&& + (\partial_\sigma X^{\rd}\gamma^{\rd}
+ \mu X^{\rdp}\gamma^{\rdp}\Omega \Pi) (S^1_- + \Omega S^2_-)
\Bigr)_{\partial\Sigma}.
\end{eqnarray}
For $OD3$-branes, $q^-_{+ \sigma} \bigg|_{\partial\Sigma}$ in Eq. \eq{dynq+sigma}
identically vanishes since $\partial_\sigma X^{\hat{r}}=
X^{\rd}=X^{\rhp}=0$ and thus 4 dynamical supersymmetries $q^-_+$ are
conserved as was shown in \ct{hikida,ggsns}.
In most cases except $OD3$-branes, $q^-_{\pm \sigma} \bigg|_{\partial\Sigma}$
appears not to vanish due to the presence of Neumann coordinates
proportional to $\mu$ such as $X^{\rd}$ and $X^{\rh}$.
However, there are some special cases satisfying
\begin{equation}\label{mod5}
    P_+ \Omega = P_+ \Pi \;\; \mbox{or} \;\; P_+ \Pi \gamma
\end{equation}
for some $OD5$-branes in Eq. \eq{2-Gamma}.
For example, these are $OD5$-branes with $\Omega =
\frac{1}{2}(\gamma^{1}\!-\!\gamma^{6})(\gamma^{2}\! \mp
\!\gamma^{5})\gamma^{34}$ and $\Omega =  \frac{1}{2}(\gamma^{1}\!-\!\gamma^{6})
(\gamma^{2}\! \mp \!\gamma^{5})\gamma^{78}$ in Table 1.
The $OD5$-brane with $\Omega =
\frac{1}{2}(\gamma^{1}\!-\!\gamma^{6})(\gamma^{2}\! \pm
\!\gamma^{5})\gamma^{37}$ in Table 1 does not satisfy Eq.
\eq{mod5}.

For the $OD5$-branes satisfying Eq. \eq{mod5},
$q^-_{\pm \sigma} \bigg|_{\partial\Sigma}$ vanish if the
following boundary conditions are true, respectively:
\bea \la{+bc-od5}
&& \partial_\sigma X^{\rd} - \mu X^{\rd} = 0, \quad \forall \rd \in PN, \\
\la{+bcn-od5}
&& \partial_\sigma X^{\hat{r}} =0, \quad \forall \rh \in ON, \\
\la{+bcd-od5}
&&  X^{\rhp} = 0, \quad \forall \rhp \in OD,
\eea
and
\bea \la{-bc-od5}
&& \partial_\sigma X^{\rh} - \mu X^{\rh} = 0, \quad \forall \rh \in ON, \\
&& \partial_\sigma X^{\rd} =0, \quad \forall \rd \in PN, \\
\la{-bcd-od5}
&&  X^{\rdp} = 0, \quad \forall \rdp \in PD.
\eea
Here abbreviations were used according to the index convention
described in footnote \ref{notation}. Note that some Neumann
boundary conditions should be modified as in the $D_+5$-brane to
preserve dynamical supersymmetry. Note also that the two kinds of
dynamical supersymmetry, $q^-_+$ and $q^-_-$, cannot
simultaneously be preserved since two sets of boundary condition,
Eqs. \eq{+bc-od5}-\eq{+bcd-od5} and Eqs.
\eq{-bc-od5}-\eq{-bcd-od5}, cannot simultaneously be compatible with each other.

One may test whether an OD-brane with a constant flux as the
$D_{-}5$-brane can preserve dynamical supersymmetry or not. The
currents $q_{\pm \sigma}^-$ in Eqs. \eq{dynq+sigma} and \eq{dynq-sigma} immediately
show that it is not possible since the condition $\gamma^r =
\pm \gamma^{r^\prime} \Omega\Pi$ is never satisfied for OD-branes.
For $OD_-5$-brane with $\Omega = - \gamma$, for example,
$\gamma^{\hat{r}} \neq \pm \gamma^{\hat{r}^\prime} \Omega\Pi$.

From the worldsheet point of view, the modification of the Neumann
boundary condition in Eq. \eq{+bc-od5} and Eq. \eq{-bc-od5} corresponds to the
addition of the following boundary terms, respectively:
\be \la{+boundary}
S_B = \frac{1}{2\pi \a^\prime p^+} \int d\tau \int_{0}^{2\pi \a^\prime |p^+|}
d\sigma F^{-\rd} \p_\sigma X^{\rd} =
\frac{\mu}{4 \pi \a^\prime p^+} \int_{\p \Sigma}  d\tau X^{\rd} X^{\rd},
\ee
where the Born-Infeld flux $F^{-\rd}$ is given by
\be \la{+bi}
F^{-\rd}= \mu X^{\rd}, \qquad \rd = \dot{3}, \dot{4},
\ee
and
\be \la{-boundary}
S_B = \frac{1}{2\pi \a^\prime p^+} \int d\tau \int_{0}^{2\pi \a^\prime |p^+|}
d\sigma F^{-\rh} \p_\sigma X^{\rh} =
\frac{\mu}{4 \pi \a^\prime p^+} \int_{\p \Sigma}  d\tau X^{\rh} X^{\rh},
\ee
where
\be \la{-bi}
F^{-\rh}= \mu X^{\rh}, \qquad \rh = \hat{1}, \hat{2}.
\ee
Thus we have to check whether the Born-Infeld fluxes such as \eq{+bi}
and \eq{-bi} are consistent with the equation of motion for a worldvolume
gauge field on the $OD5$-brane. We will show the details in
Appendix A that a worldvolume Chern-Simons coupling does give rise
to the correct boundary conditions \eq{+bc-od5} and \eq{-bc-od5}.

Let us take a more close look at the $OD5$-brane with $\Omega =
\frac{1}{2}(\gamma^{1}\!-\!\gamma^{6})(\gamma^{2}\! \pm
\!\gamma^{5})\gamma^{37}$ as well as the $OD7$-brane with
$\frac{1}{2}(\gamma^{1}\!-\!\gamma^{6})(\gamma^{2}\! \pm \!\gamma^{5})\gamma^{3478}$
in Table 1. If the boundary conditions in Eqs.
\eq{+bcn-od5}-\eq{+bcd-od5} hold, $q^-_{+\sigma}$ at the worldsheet boundary
$\partial\Sigma$ is involved only with $S_+^A$ spinors whose
all eigenvalues of the matrix $\Gamma = \Omega \Pi \Omega \Pi$ are
1. Thus we meet a similar situation to the case of
$(+,-,n,n)$-brane. Indeed it turns out that some conserved
dynamical supersymmetries exist in this case too. To see this, let
us decompose the Neumann coordinates in the similar way as $\rd=(\rd_1,
\rd_2)$. For the $OD5$-brane, for example, $\rd_1 = 3, \rd_2=7$.
For the projection matrix \eq{proj-d+}, we get
\be \label{od37-comm}
[P_\pm^{D_+},\gamma^{\rd_1}]=0, \qquad
P_\pm^{D_+} \gamma^{\rd_2}=\gamma^{\rd_2}P_\mp^{D_+}.
\ee

Now define doubly projected supercharges
\be \la{d537-susy}
q^-_{+\pm} \equiv P_\pm^{D_+} P_+ (Q^{-1}- \Omega Q^{-2}).
\ee
Then, from Eq. \eq{dynq+sigma}, we get $q^-_{+ \pm \sigma}$ at the
worldsheet boundary $\p \Sigma$
\begin{eqnarray}\label{37q-sigma}
q^-_{+ \pm \sigma} \bigg|_{\partial\Sigma}&=& - \sqrt{\frac{1}{2p^+}}
\Bigl( \partial_\sigma X^{\rd_1} \gamma^{\rd_1}
\mp \mu X^{\rd_1} \gamma^{\rd_1}) P_\pm^{D_+}(S^1_+ + \Omega S^2_+) \xx
&& + (\partial_\sigma X^{\rd_2} \gamma^{\rd_2}
\pm \mu X^{\rd_2} \gamma^{\rd_2}) P_\mp^{D_+}(S^1_+ + \Omega S^2_+)
\Bigr)_{\partial\Sigma}.
\end{eqnarray}
We see that two dynamical supersymmetries are preserved only if
the boundary conditions Eqs. \eq{+bcn-od5}-\eq{+bcd-od5} and
the following Neumann boundary conditions given by
\be \la{mbc-37}
\Bigl(\partial_\sigma X^{\rd_1}
\mp \mu X^{\rd_1} \Bigr)_{\partial\Sigma} = 0 =
\Bigl(\partial_\sigma X^{\rd_2}
\pm \mu X^{\rd_2}\Bigr)_{\partial\Sigma}
\ee
hold. The boundary condition \eq{mbc-37} can also be realized by introducing
a gauge field excitation of the similar form as Eq. \eq{d+boundary}
with the Born-Infeld flux $F^{-\rd}$ satisfying
\be \la{37bi}
\p_{\rd} F^{-\rd}= 0, \qquad \rd = (\rd_1,\rd_2).
\ee

\begin{table}[tbp]
\begin{center}
\begin{tabular}{|c|c|c|c|c|c|} \hline
D-brane type & $\Gamma$         & $\Omega$             & $q^+_{D_-}$ & $q^+_{D_+}$ & $q^-$\\
\hline
$\mathbf{D_{-}5}$      & $ -1 $    & $(3,1), \; (1,3)$   &   8  & 0  &  8    \\
\hline
$\mathbf{D_{+}(2n+1)}$      & $ 1 $    & $(n,n), \; \; n=1,2,3,4$   & 0 &  8   & 4    \\
\hline
$OD3$        & $\pm \gamma^{1256}$ & $\frac{1}{2}(\gamma^{1}\!-\!\gamma^{6})
(\gamma^{2}\! \pm \!\gamma^{5})$     & 4 & 4  &  4 \\
\hline
$\mathbf{OD5}$       & $\pm \gamma^{1256}$ & $\frac{1}{2}(\gamma^{1}\!-\!\gamma^{6})
             (\gamma^{2}\! \mp \!\gamma^{5})\gamma^{34}$    & 4 & 4  & 4  \\
             &                 & $\frac{1}{2}(\gamma^{1}\!-\!\gamma^{6})
(\gamma^{2}\! \mp \!\gamma^{5}) \gamma^{78}$          &  &  &    \\
\hline
$\mathbf{OD5} $ & $\pm \gamma^{1256}$ & $\frac{1}{2}(\gamma^{1}\!-\!\gamma^{6})
             (\gamma^{2}\! \pm \!\gamma^{5})\gamma^{37}$  &  4 & 4 & 2  \\
\hline
$\mathbf{OD7}$ & $\pm \gamma^{1256}$ & $\frac{1}{2}(\gamma^{1}\!-\!\gamma^{6})
             (\gamma^{2}\! \pm \!\gamma^{5})\gamma^{3478}$  & 4 & 4  & 2 \\
\hline
$OD_{+}5$  & $ \gamma $    & $\frac{1}{4}(\gamma^{1}\!-\!\gamma^{6})
(\gamma^{2}\! + \!\gamma^{5}) (\gamma^{3}\!-\!\gamma^{8})(\gamma^{4}\!+\!\gamma^{7})$
&  0 & 8 & 0  \\
\hline
$OD_{-}5$  & $ -\gamma $    & $\frac{1}{4}(\gamma^{1}\!-\!\gamma^{6})
(\gamma^{2}\! - \!\gamma^{5}) (\gamma^{3}\!-\!\gamma^{8})(\gamma^{4}\!+\!\gamma^{7})$
& 8  & 0 & 8  \\
\hline
\end{tabular}
\end{center}
\caption{Supersymmetry of flat D-branes.
A D-brane with a gauge field condensate is denoted by the
boldface. $q^+_{D_-} (q^+_{D_+})$ is the number of unbroken
kinematical supersymmetry of $D_-$-type ($D_+$-type).}
\label{tabletwo}
\end{table}

The results on the dynamical supersymmetry preserved by D-branes
in Table 1 have been summarized in Table 2, in which we omitted
BPS $D_\pm$-branes preserving 16 supersymmetries without flux since they
have already been identified in \ct{kim}.

\section{Supersymmetry of Intersecting D-branes}

We now analyze the supersymmetry of intersecting D-branes for
which open string mode expansion has been given in Sec. 4. In
order to put our discussion on a general ground,
we will not assume anything about the matrix $\Gamma$,
so we are including the D-branes in \eq{1-Gamma} and \eq{3-Gamma} as well.
The supersymmetry of intersecting $D_\pm$-branes was completely
identified in \ct{cha} using the Green-Schwarz worldsheet formulation
which can also be applied to more general class of D-branes under
consideration. In what follows, all supercharges are assumed to be
expressed in view of the D-brane at $\sigma = 0$ in the same way as
the mode expansion in Sec. 4.

In general, the unbroken supersymmetry of intersecting D-branes is
the `intersection' of supersymmetries preserved by each brane. The
intersection is characterized by the projection matrices $I_\pm$
in Eq. \eq{i+-}. In Sec. 5, we showed that
the conserved supersymmetry of a single D-brane is described by
two kinds of projection matrices, $P_\pm$ and $P_\pm^{D_+}$, in
Eqs. \eq{projection} and \eq{proj-d+}. Thus the unbroken supersymmetry
of intersecting D-branes shall be completely characterized by
these three kinds of projection matrices. From the definitions of
the projection matrices, one can see that they are mutually
commuting, viz.,
\be \la{3-proj-comm}
[I_\pm, P_\pm] = [I_\pm, P_\pm^{D_+}] = [P_\pm, P_\pm^{D_+}] = 0,
\ee
if the condition \eq{pi-Gamma} holds. From now on, we will assume
it.

For an intersection of half BPS D-branes, it was shown in \ct{cha}
that the supersymmetry of intersecting D-branes is given by
$I_\pm q^+$ and $I_\pm q^-$ where the kinematical supercharge $q^+$
is defined by Eq. \eq{ks-} and \eq{ks+} and the dynamical supercharge
$q^-$ is by Eq. \eq{dyn-susy}. For the present problem, however,
the supersymmetry of a single D-brane is in general a subset of
$q^+$ and $q^-$, represented by $P_\pm$ and $P_\pm^{D_+}$,
as illustrated in Table 2. Nevertheless, since all the projection
matrices mutually commute as in Eq. \eq{3-proj-comm}, we first find
the condition for the supercharges $I_\pm q^+$ and $I_\pm q^-$
to be conserved and then construct unbroken supersymmetries using
the projection matrices $P_\pm$ and $P_\pm^{D_+}$.
This immediately implies that
intersecting D-branes preserve no supersymmetry when $\Gamma^2_0 = - 1$
since $q^+$ and $q^-$ were originally not conserved quantities.

Following the recipe explained above, it is simple to check from
the conservation laws \eq{con-law-kin-} and \eq{con-law-kin+} using the
identities \eq{i+-pi}, \eq{identity-a}, and
\eq{identity-b} that the supercharge
\be \la{inter-kin}
q^+ = I_\pm ( q^+_{D_-} + q^+_{D_+} )
\ee
is conserved for A-type branes with $I_+$ and for B-type branes
with $I_-$ only if $\Gamma^2_0 = 1$. Here $q^+_{D_-}$ and $q^+_{D_+}$ are defined
by Eq. \eq{ks-d-} and Eq. \eq{ks-d+}, respectively. Since we only
required the condition \eq{pi-Gamma} for the matrices $ \Gamma_0$ and $\Gamma_\pi$,
there are two kinds of intersection to preserve the kinematical
supersymmetry:
\bea \la{inter-same}
&& \Gamma_0 = \Gamma_\pi, \\
\la{inter-diff}
&& \Gamma_0 = \Gamma_\pi \gamma.
\eea
The special case, $\Gamma_0 = \pm 1 = \Gamma_\pi$, in Eq.
\eq{inter-same} was already analyzed in \ct{cha} and their
unbroken supersymmetry was completely identified.
Eq. \eq{inter-diff} implies that there are
supersymmetric intersections between $D_\pm$-branes and
$OD5_\pm$-branes as well as between $ODp$-branes with, e.g.,
$\Gamma_0 = \pm \gamma^{1256}$ and $ODq$-branes with
$\Gamma_\pi = \pm \gamma^{3478}$.

For a given class in \eq{inter-same} or \eq{inter-diff},
the number of unbroken kinematical supersymmetry depends on the
total number of ND and DN directions which is captured by the
matrix in \eq{omega-sol}. Since a single D-brane originally had
$8 = 4_{D_-} + 4_{D_+}$ kinematical supersymmetries as shown in
Table 2, intersecting D-branes, depending on the total number of
ND directions, will have $8 = 4_{D_-} + 4_{D_+}$, $4 = 2_{D_-} +
2_{D_+}$, or 0 kinematical supersymmetries according to Eq. \eq{inter-kin}.

Now we will discuss the dynamical supersymmetry of intersecting
D-branes. It is useful to recall the (anti-)commutation relations
between
$\gamma^I = \{\gamma^r, \gamma^{r^\prime}, \gamma^i,
\gamma^{i^\prime} \}, \; \Omega_0$ and $\Omega_\pi$
to find conserved dynamical supersymmetries:
\bea \la{gamma-omega-comm}
&& \{\gamma^r, \Omega_0 \}= \{\gamma^i, \Omega_0 \}
=[\gamma^{r^\prime}, \Omega_0 ]= [ \gamma^{i^\prime}, \Omega_0 ]=0, \\
\la{gamma-inik}
&& \{\gamma^r, \Omega_\pi \}= \{\gamma^{i^\prime}, \Omega_\pi \}
=[\gamma^{r^\prime}, \Omega_\pi ]= [ \gamma^i, \Omega_\pi ]=0.
\eea
Here we adopted the indices explained in footnote
\ref{inter-index}.

To follow the same strategy explained above, we introduce a
dynamical supercharge of intersecting D-branes defined
by\footnote{\la{diff-q} Note that there is a minor difference
between Eq. \eq{inter-dyn-susy} and that
in \ct{cha} due to the different definition about $q^-$ mentioned
in footnote \ref{diff-def}.}
\be \la{inter-dyn-susy}
q^- = I_\pm (Q^{-1}- \Omega Q^{-2})
\ee
where $I_+$ is for A-type branes and $I_-$ for B-type branes.
Since the supercharge in Eq. \eq{inter-dyn-susy} is just the
projection of that in Eq. \eq{dyn-susy}, it is obvious that
the supercharge in Eq. \eq{inter-dyn-susy}
also satisfies the conservation law \eq {con-law-dyn}.
Using the (anti-)commutation relations \eq{gamma-omega-comm} and
\eq{gamma-inik}, it is easy to find $q^-_\sigma$ from Eq. \eq{ds-current}
which is given by
\bea \la{inter-ds-current}
q^-_\sigma &=& \sqrt{\frac{1}{2p^+}}
\Bigl( (\p_\tau X^r \gamma^r - \p_\sigma X^{r^\prime}
\gamma^{r^\prime}) I_{\pm}(S^1 - \Omega_0 S^2)
+ (\p_\tau X^i \gamma^i - \p_\sigma X^{i^\prime}
\gamma^{i^\prime}) I_{\mp}(S^1 - \Omega_0 S^2) \xx
&& + (\p_\tau X^{r^\prime} \gamma^{r^\prime} - \p_\sigma X^r \gamma^r)
I_\pm (S^1 + \Omega_0 S^2) +
(\p_\tau X^{i^\prime} \gamma^{i^\prime} - \p_\sigma X^i \gamma^i)
I_\mp (S^1 + \Omega_0 S^2) \xx
&& + \mu X^r \gamma^r \Omega_0 \Pi I_\pm  (S^1 + \Gamma_0 \Omega_0 S^2)
+ \mu X^i \gamma^i \Omega_0 \Pi I_\mp  (S^1 + \Gamma_0 \Omega_0 S^2) \xx
&& - \mu X^{r^\prime} \gamma^{r^\prime} \Omega_0 \Pi I_\pm (S^1 -
\Gamma_0 \Omega_0  S^2) - \mu X^{i^\prime} \gamma^{i^\prime} \Omega_0 \Pi I_\mp
(S^1 - \Gamma_0 \Omega_0  S^2) \Bigr).
\eea

Since the case $\Gamma_0^2 = -1$ cannot preserve any
supersymmetry, we will focus on the case $\Gamma_0^2 = 1$. To find
the condition for $q^-_\sigma |_{\p \Sigma}$ to vanish, we first
decompose the supercharge $q^-$ in Eq. \eq{inter-dyn-susy} in
terms of the projection matrix $P_\pm = \half (1\pm \Gamma_0)$:
\be \la{inter-dyn-susy-pro1}
q^-_\pm \equiv P_\pm I_\pm (Q^{-1}- \Omega Q^{-2}).
\ee
Using Eqs. \eq{p+-comm1}-\eq{p+-comm2} and adopting the notation
in footnote \ref{inter-index}, it is then easy to see that
\bea \la{inter-ds-current-1+}
q^-_{+\sigma} |_{\p \Sigma} &=& - \sqrt{\frac{1}{2p^+}}
\Bigl( (\p_\sigma X^{\rd} \gamma^{\rd} - \mu X^{\rd} \gamma^{\rd} \Omega_0 \Pi )
I_{\pm}(S^1_+ + \Omega_0 S^2_+) \xx
&& + (\p_\sigma X^{\rh} \gamma^{\rh} + \mu X^{\rhp} \gamma^{\rhp} \Omega_0 \Pi )
I_{\pm}(S^1_- + \Omega_0 S^2_-) \xx
&& + (\p_\sigma X^{\id} \gamma^{\id} - \mu X^{\id} \gamma^{\id} \Omega_0 \Pi )
I_\mp (S^1_+ + \Omega_0 S^2_+) \xx
&& + (\p_\sigma X^{\ih} \gamma^{\ih} + \mu X^{\ihp} \gamma^{\ihp} \Omega_0 \Pi )
I_\mp (S^1_- + \Omega_0 S^2_-) \xx
&& + (\p_\sigma X^{\idp} \gamma^{\idp}
+ \mu X^{\idp} \gamma^{\idp} \Omega_0 \Pi )
I_\mp (S^1_+ - \Omega_0 S^2_+) \xx
&& + (\p_\sigma X^{\ihp} \gamma^{\ihp}
- \mu X^{\ih} \gamma^{\ih} \Omega_0 \Pi )
I_\mp (S^1_- - \Omega_0 S^2_-) \Bigr)
\eea
and
\bea \la{inter-ds-current-1-}
q^-_{-\sigma} |_{\p \Sigma} &=& - \sqrt{\frac{1}{2p^+}}
\Bigl( (\p_\sigma X^{\rd} \gamma^{\rd} + \mu X^{\rdp} \gamma^{\rdp} \Omega_0 \Pi )
I_{\pm}(S^1_- + \Omega_0 S^2_-) \xx
&& + (\p_\sigma X^{\rh} \gamma^{\rh} - \mu X^{\rh} \gamma^{\rh} \Omega_0 \Pi )
I_{\pm}(S^1_+ + \Omega_0 S^2_+) \xx
&& + (\p_\sigma X^{\idp} \gamma^{\idp} - \mu X^{\id} \gamma^{\id} \Omega_0 \Pi )
I_\mp (S^1_- - \Omega_0 S^2_-) \xx
&& + (\p_\sigma X^{\ihp} \gamma^{\ihp} + \mu X^{\ihp} \gamma^{\ihp} \Omega_0 \Pi )
I_\mp (S^1_+ - \Omega_0 S^2_+) \xx
&& + (\p_\sigma X^{\id} \gamma^{\id}
+ \mu X^{\idp} \gamma^{\idp} \Omega_0 \Pi )
I_\mp (S^1_-  +  \Omega_0 S^2_-) \xx
&& + (\p_\sigma X^{\ih} \gamma^{\ih}
- \mu X^{\ih} \gamma^{\ih} \Omega_0 \Pi )
I_\mp (S^1_+ + \Omega_0 S^2_+) \Bigr)
\eea
where we dropped obviously vanishing terms at the boundary
${\p \Sigma}$.
The second identities in Eqs. \eq{identity-a}-\eq{identity-b} played
a crucial role to derive the above result.

It is straightforward to find intersecting D-brane configurations
preserving some dynamical supersymmetry from Eqs.
\eq{inter-ds-current-1+}-\eq{inter-ds-current-1-}.
Here the second identities in Eqs. \eq{identity-a}-\eq{identity-b} will play
a crucial role again to do this job. Also recall the following
facts
\be \la{chiral-eigen}
\gamma Q^{-A} = - Q^{-A}, \qquad  \gamma S^A  = S^A.
\ee

$\bullet \; \Gamma_0 = \Gamma_\pi = - 1$: This case is exactly the
$D_- - D_-$ brane intersections studied in \ct{cha}. The worldsheet
current is given by Eq. \eq{inter-ds-current-1-}
where only $S^A_- = S^A$ spinors survive. From Eq.
\eq{inter-ds-current}, one can see that there can be
supersymmetric intersections of $D_-5$-branes with a constant flux
when $\sharp_{ND}=0,4,8$. For example, let us consider a
$\sharp_{ND}=4$ intersection of $D5$-brane extended along
$(1,2,3,5)$ directions with flux $F_{+5}= \mu q_1$ and $D5$-brane along
$(1,2,4,6)$ directions with flux $F_{+6}= \mu q_2$. In this case,
we have $\Omega_0 \Pi = - \gamma^{45}, \; \Omega_\pi \Pi =
\gamma^{36}$, and $\Omega_0^T \Omega_\pi = \gamma^{3456}$. It is
easy to check that $q^-_{\sigma} |_{\p \Sigma} = 0$ if the
following boundary conditions are satisfied
\bea \la{intsect-d5-flux}
&& \p_\sigma X^{1,2,3}|_{\sigma=0} = 0 = X^{6,7,8}|_{\sigma=0},
\quad (\p_\sigma X^5 - \mu q_1)_{\sigma=0} = 0 =
(X^4-q_1)_{\sigma=0}, \xx
&& \p_\sigma X^{1,2,4}|_{\sigma=\pi} = 0 = X^{5,7,8}|_{\sigma=\pi},
\quad (\p_\sigma X^6 - \mu q_2)_{\sigma=\pi} = 0 =
(X^3-q_2)_{\sigma=\pi}.
\eea
Similarly, for a $\sharp_{ND}=8$ intersection of $D5$-brane extended along
$(1,2,3,5)$ directions with flux $F_{+5}= \mu q_1$ and $D5$-brane along
$(4,6,7,8)$ directions with flux $F_{+4}= \mu q_2$, one can also check
that $q^-_{\sigma} |_{\p \Sigma} = 0$.

$\bullet \; \Gamma_0 = \Gamma_\pi =  1$: This case corresponds to the
$D_+ - D_+$ brane intersections. The worldsheet current is given by
Eq. \eq{inter-ds-current-1+} where only $S^A_+ = S^A$ spinors survive.
It is convenient to decompose the supercharge $q^-_+$ in
terms of the projection matrix $P_\pm^{D_+} = \half ( 1 \pm \Omega_0 \Pi)$
as in Eq. \eq{proj-d+susy}. Note that the third term
in Eq. \eq{inter-ds-current-1+} gives a
constraint only at $\sigma=0$, i.e., a Neumann boundary condition
while the fifth term does only at $\sigma=\pi \a$, also a Neumann boundary
condition. Using Eqs. \eq{d+p+-comm1}-\eq{d+p+-comm2}, we get
\bea \la{bc-d+d+1}
&& (\p_\sigma X^{r_1} \mp \mu X^{r_1})_{\p \Sigma} = 0,
\qquad (\p_\sigma X^{r_2} \pm \mu X^{r_2})_{\p \Sigma} = 0, \\
\la{bc-d+d+2}
&& (\p_\sigma X^{i_1} \mp \mu X^{i_1})_{\sigma = 0} = 0,
\qquad (\p_\sigma X^{i_2} \pm \mu X^{i_2})_{\sigma = 0} = 0, \\
\la{bc-d+d+3}
&& (\p_\sigma X^{i^\prime_1} \mp \mu X^{i^\prime_1})_{\sigma = \pi \a} = 0,
\qquad (\p_\sigma X^{i^\prime_2}
\pm \mu X^{i^\prime_2})_{\sigma = \pi \a} = 0,
\eea
where the upper sign is for $P_+^{D_+}$ and the lower one for
$P_-^{D_+}$. We see that the $D_+ - D_+$ brane intersections can
preserve some dynamical supersymmetries as long as the worldvolume
fluxes are consistently aligned as the
way in Eqs. \eq{bc-d+d+1}-\eq{bc-d+d+3}. Consider, for example,
$(+,-,4,0)-(+,-,1,1)$ brane intersection. The supersymmetric
boundary conditions must be
\be \la{40-11}
\p_\sigma X^{r_1} - \mu X^{r_1} =0, \quad \p_\sigma X^{i_1} - \mu
X^{i_1}= 0, \quad \p_\sigma X^{i_2^\prime} + \mu X^{i_2^\prime} =
0
\ee
for $r_1 =1, \; i_1 =2,3,4$ and $i_2^{\prime} =5$, for example.
However, $(+,-,4,0)-(+,-,0,4)$ brane intersection cannot satisfy Eq. \eq{bc-d+d+2}
and Eq. \eq{bc-d+d+3} simultaneously and thus preserves no dynamical supersymmetry
as was shown in \ct{cha}.

$\bullet \; \Gamma_0 = \Gamma_\pi = - \gamma$: This case is the
$OD_-5 - OD_-5$ brane intersection. The worldsheet current is
given by Eq. \eq{inter-ds-current-1+} where only $S^A_- = S^A$ spinors survive
since Eq. \eq{chiral-eigen}. The dynamical supersymmetry can be
preserved as long as the two branes are placed at origin, viz.,
\be \la{od5-od5}
X^{\rhp}|_{\p \Sigma} = X^{\ihp}|_{\sigma = 0}
= X^{\ih}|_{\sigma = \pi \a} = 0,
\ee
and they satisfy the usual Neumann boundary conditions. These
boundary conditions are the same as the $D_-5 - D_-5$ brane case
as expected.

$\bullet \; \Gamma_0 = \Gamma_\pi = \gamma$: This case is the
$OD_+5 - OD_+5$ brane intersection. The worldsheet current is
given by Eq. \eq{inter-ds-current-1-} where only $S^A_+ = S^A$ spinors survive
since Eq. \eq{chiral-eigen}. Any dynamical supersymmetry is not preserved
as expected.

$\bullet \; \Gamma_0 = \Gamma_\pi = \pm \gamma^{1256}$: First, note that
for $\Omega_0 = \frac{1}{2}(\gamma^{1}\!-\!\gamma^{6})(\gamma^{2}\! \mp
\!\gamma^{5})\gamma^{34}$ in Table 1
\be \la{omega-pi1}
\Omega_0 \Pi = P_+ + \gamma^{16} P_-
\ee
and for $\Omega_0 =  \frac{1}{2}(\gamma^{1}\!-\!\gamma^{6})
(\gamma^{2}\! \mp \!\gamma^{5})\gamma^{78}$
\be \la{omega-pi2}
\Omega_0 \Pi = \mp (P_+ - \gamma^{16} P_-) \gamma.
\ee
For the first case \eq{omega-pi1}, there are two ways to preserve
dynamical supersymmetries. One set can be read off
from Eq. \eq{inter-ds-current-1+}:
\bea \la{1-34}
&& X^{\rhp}|_{\p \Sigma} = X^{\ihp}|_{\sigma = 0}
= X^{\ih}|_{\sigma = \pi \a} = 0, \xx
&& \p_\sigma X^{\rh}|_{\p \Sigma} = \p_\sigma X^{\ih}|_{\sigma = 0}
= \p_\sigma X^{\ihp}|_{\sigma = \pi \a} = 0, \\
&& (\p_\sigma X^{\rd} - \mu X^{\rd})_{\p \Sigma}
= (\p_\sigma X^{\id} - \mu X^{\id})_{\sigma = 0}
= (\p_\sigma X^{\idp} + \mu X^{\idp})_{\sigma = \pi \a}= 0. \nonumber
\eea
The other set can be done from Eq. \eq{inter-ds-current-1-}:
\bea \la{2-34}
&& X^{\rdp}|_{\p \Sigma} = X^{\idp}|_{\sigma = 0}
= X^{\id}|_{\sigma = \pi \a} = 0, \xx
&& \p_\sigma X^{\rd}|_{\p \Sigma} = \p_\sigma X^{\id}|_{\sigma = 0}
= \p_\sigma X^{\idp}|_{\sigma = \pi \a} = 0, \\
&& (\p_\sigma X^{\rh} - \mu X^{\rh})_{\p \Sigma}
= (\p_\sigma X^{\ih} - \mu X^{\ih})_{\sigma = 0}
= (\p_\sigma X^{\ihp} + \mu X^{\ihp})_{\sigma = \pi \a}= 0. \nonumber
\eea
For the second case \eq{omega-pi2}, the result is similar.

For the $OD5$-brane with $\Omega_0 =
\frac{1}{2}(\gamma^{1}\!-\!\gamma^{6})(\gamma^{2}\! \pm
\!\gamma^{5})\gamma^{37}$ and the $OD7$-brane with
$\Omega_0 = \frac{1}{2}(\gamma^{1}\!-\!\gamma^{6})
(\gamma^{2}\! \pm \!\gamma^{5})\gamma^{3478}$
in Table 1, we can apply the same method as Sec. 5. After defining
(now triply) projected supercharges like as \eq{d537-susy}, one can easily find
the supersymmetric boundary conditions. These are the
same as Eqs. \eq{1-34}-\eq{2-34} except that the Neumann
boundary conditions modified by gauge field condensates are split
into two $SO(4)$ directions in a way consistent with the
equation of motion such as Eq. \eq{37bi}.

There is a subtle point when we consider an intersection of the
$OD5$-brane with $\Omega_0 = \frac{1}{2}(\gamma^{1}\!-\!\gamma^{6})
(\gamma^{2}\! \pm \!\gamma^{5})\gamma^{34}$ and the $OD5$-brane
with $\Omega_\pi = \frac{1}{2}(\gamma^{1}\!-\!\gamma^{6})(\gamma^{2}\! \pm
\!\gamma^{5})\gamma^{37}$ as a typical example. If we choose the
former brane as a reference brane, the projection matrix
\eq{proj-d+} is no longer needed to construct a conserved
supercharge. However, if we choose the latter one, the projection
matrix, at first sight, seems to be necessary to construct a
conserved supercharge. This seems to introduce a contradictory
result that the conserved dynamical supersymmetry depends on our
choice of a reference brane. This puzzle can be resolved as
follows. Suppose that we choose the $OD5$-brane described by
$\Omega_\pi$ as a reference brane.
Since we have already had the projection matrix $I_\pm$,
the gluing matrix $\Omega_\pi$ can be represented by $\Omega_0$ as follows
\be \la{pi-0}
\Omega_\pi = \Omega_0 (I_+ - I_-).
\ee
Thus it is not necessary to further introduce the projection
matrices \eq{proj-d+} and the dynamical supersymmetry is not
further reduced in the case at hand. The other cases, e.g. the
$OD7$-branes with $\Omega_\pi = \frac{1}{2}(\gamma^{1}\!-\!\gamma^{6})
(\gamma^{2}\! \pm \!\gamma^{5})\gamma^{3478}$, are similar too.

$\bullet \; \Gamma_0 = -1, \; \Gamma_\pi = - \gamma$: This case is the
$D_-p - OD_-5$ brane intersection. However any dynamical
supersymmetry cannot be preserved. The reason is simple: If we
choose $P_\pm = \half(1\pm \Gamma_0)$, the non-vanishing
supercharge is $q^-_-$. As it should be, the physics must be
equally well described in view of the D-brane at $\sigma = \pi \a$
as well where $P_\pm = \half(1\pm \Gamma_\pi)$. Then the non-vanishing
supercharge is now $q^-_+$. But these two charges cannot
simultaneously be compatible with each other
as we discussed below Eq. \eq{-bcd-od5}.
This kind of thing does not happen for kinematical supersymmetry
since $\gamma Q^{+A} = Q^{+A}$, which is a reason why it can be
preserved in the case of $\Gamma_0 = \Gamma_\pi \gamma$.

$\bullet \; \Gamma_0 = 1, \; \Gamma_\pi = \gamma$: This case is the
$D_+p - OD_+5$ brane intersection. Similarly this case has no
dynamical supersymmetry.

$\bullet \; \Gamma_0 = \pm \gamma^{1256}, \;
\Gamma_\pi = \pm \gamma^{3478}$: This case corresponds to the
$ODp - ODq$ brane intersection. This case has no dynamical
supersymmetry either since $(1 \pm \gamma^{1256})(1 \pm \gamma^{3478})
=(1 \pm \gamma^{1256})(1 + \gamma)$.

\begin{table}[tbp]
\begin{center}
\begin{tabular}{|c|c|c|c|} \hline
D-brane type &   Intersection           &  $q^+$         & $q^-$  \\
\hline
$\Gamma_0 = \Gamma_\pi = -1$      & $D_-p - D_-q $    & $\nu$
& $\half \Tr (1-\gamma)I_\pm$  \\
\cline{2-4}
    & $\mathbf{D}_- 5 - \mathbf{D}_- 5 $    & $\nu$
& $\half \Tr (1-\gamma) I_\pm $ \\
\hline
$\Gamma_0 = \Gamma_\pi =1$      & $D_+ 1 - \mathbf{D}_+ p $    & $\nu$
& $\half \Tr (1-\gamma)P_\pm^{D_+} I_\pm $\\
\cline{2-4}
    & $\mathbf{D}_+ p - \mathbf{D}_+ q $    & $\nu$
& $\half \Tr (1-\gamma)P_\pm^{D_+} I_\pm $\\
\hline
$\Gamma_0 = \Gamma_\pi =-\gamma$      & $OD_- 5 - OD_-5 $    & $\nu$
& $\half \Tr (1-\gamma)I_\pm$ \\
\hline
$\Gamma_0 = \Gamma_\pi =\gamma$      & $OD_+ 5 - OD_+5 $    & $\nu$
& 0 \\
\hline
$\Gamma_0 = \Gamma_\pi =\pm \gamma^{1256}$      & $\mathbf{OD}p - \mathbf{OD}q$
& $\nu$ & $\half \Tr (1-\gamma)P_\pm^{D_+} P_\pm I_\pm $\\
\hline
$\Gamma_0 = - 1,  \Gamma_\pi = - \gamma$      & $D_-p - OD_-5 $    & $\nu$
& 0 \\
\hline
$\Gamma_0 = 1,  \Gamma_\pi = \gamma$      & $D_+p - OD_+5 $    & $\nu$
& 0 \\
\hline
$\Gamma_0 =\pm \gamma^{1256}, \Gamma_\pi =\pm \gamma^{3478}$   & $ODp - ODq $
& $\nu$ & 0 \\
\hline

\end{tabular}
\end{center}
\caption{Supersymmetry of intersecting D-branes. $\nu= n_{D_-} + n_{D_+}$
is the number of unbroken kinematical supersymmetry where $n_{D_-} = \half \Tr
(1+\gamma) P_- I_\pm$ ($D_-$-type) and $n_{D_+} = \half \Tr
(1+\gamma) P_+ I_\pm$ ($D_+$-type). A D-brane with a gauge field condensate
is denoted by the boldface.}
\label{tablethree}
\end{table}

We summarized the supersymmetry preserved by various
configurations of intersecting D-branes in Table 3. The number of
each type of kinematical supersymmetries depends on the total
number of ND and DN directions in a way determined by the
projection matrix $I_\pm P_-$ ($D_-$-type) or $I_\pm P_+$ ($D_+$-type)
as was shown in Eq. \eq{inter-kin}. The number
of dynamical supersymmetries listed in Table 3 may be counted
using the relations \eq{omega-pi1}, \eq{omega-pi2}, and \eq{pi-0}.

\section{Supersymmetric Curved D-branes}

It was shown in \ct{hikida,ggsns} that the plane wave background
\eq{pp-metric} admits supersymmetric curved D-branes as well as
oblique D-branes. We will also classify the supersymmeric curved
D-branes including those not mentioned in the previous
literatures.

To discuss curved D-branes, it is convenient to introduce the
complex coordinates as
\be \la{comp-z}
Z^i = X^i + i X^{i+4}, \qquad i=1,\cdots,4
\ee
and the fermionic creation and annihilation operators
\bea \la{b-b*}
&& b_i = \half(\gamma^i - i \gamma^{i+4}), \qquad b_i^\dagger
= \half(\gamma^i + i \gamma^{i+4}), \\
\la{anti-comm-bb*}
&& \{b_i, b_j^\dagger \}= \delta_{ij}, \qquad i,j=1,\cdots,4.
\eea
Using a general $\CN = (2,2)$ supersymmetric worldsheet theory,
Hikida and Yamaguchi showed in \ct{hikida} that a supersymmetric
curved D-brane is wrapped on a complex submanifold and the
superpotential due to the geometry \eq{pp-metric} should be
constant on its worldvolume. The maximally supersymmetric plane
wave \eq{pp-metric} has a special superpotential
\be \la{superpotential-w}
W = i(Z_1^2 + Z_2^2 + Z_3^3 + Z_4^2).
\ee

Thus possible supersymmetric curved D-branes in the background
\eq{pp-metric} can be exhausted as follows:
\bea \la{d7-curved-brane}
 D7: && Z_1^2 + Z_2^2 + Z_3^3 + Z_4^2=c_1, \\
 \la{d5-curved-brane1}
 D5: && Z_1^2 + Z_2^2 + Z_3^3 = c_1, \quad Z_4 = c_2, \\
 \la{d5-curved-brane2}
     && Z_1^2 + Z_2^2=c_1, \quad Z_3^2 + Z_4^2 = c_2, \\
 \la{d5-curved-brane3}
     && Z_1^2 + Z_2^2=c_1, \quad Z_3 = i Z_4, \\
 \la{d3-curved-brane}
 D3: && Z_1^2 + Z_2^2=c_1, \quad Z_3 = c_2, \quad Z_4 = c_3,
\eea
where $c_i$'s are complex constants. Following the Ref.
\ct{ggsns}, we identify the gluing matrices $\Omega$ for the above curved
D-branes as follows:
\bea \la{d7-curved-omega}
 D7: && \Omega = -i[b_i, b_j^\dagger]
 \frac{\bar{Z}^i Z^j}{\sum_{k=1}^4 |Z_k|^2}, \quad i,j=1,\cdots,4, \\
 \la{d5-curved-omega1}
 D5: && \Omega = -i[b_i, b_j^\dagger]\gamma^{48}
 \frac{\bar{Z}^i Z^j}{\sum_{k=1}^3 |Z_k|^2},
 \quad i,j=1,2,3, \\
 \la{d5-curved-omega2}
     && \Omega = -[b_i, b_j^\dagger][b_k, b_l^\dagger]
 \frac{\bar{Z}^i Z^j}{|Z_1|^2+|Z_2|^2}
 \frac{\bar{Z}^k Z^l}{|Z_3|^2+|Z_4|^2},
 \; i,j=1,2, \; k,l=3,4, \\
 \la{d5-curved-omega3}
 && \Omega = -\frac{i}{2}[b_i, b_j^\dagger]
 (\gamma^3+ \gamma^8)(\gamma^4 - \gamma^7)
 \frac{\bar{Z}^i Z^j}{|Z_1|^2+|Z_2|^2},
 \;\; i,j=1,2, \\
 \la{d3-curved-omega}
 D3: && \Omega = -i[b_i, b_j^\dagger]\gamma^{3478}
 \frac{\bar{Z}^i Z^j}{|Z_1|^2+|Z_2|^2},
 \qquad i,j=1,2,
\eea
where $\Omega$'s are now taken to be the product of the
$\gamma$-matrices associated to the Dirichlet directions.
The $D7$-brane in Eq. \eq{d7-curved-brane} was previously discussed in
\ct{hikida,ggsns} and the $D5$-brane in Eq. \eq{d5-curved-brane1} was in
\ct{ggsns}. Note that $\Omega^T \Omega = 1$ always and $\Omega^2= -1$
for $D7$- and $D3$-branes while $\Omega^2= 1$ for $D5$-branes.

The fermionic boundary condition is still given by Eq. \eq{bc-fermion},
but with the gluing matrix $\Omega$ depending on the nontrivial
worldvolume geometry of D-branes as shown
in Eqs. \eq{d7-curved-omega}-\eq{d3-curved-omega}.
Furthermore, it turned out \ct{ggsns} that an open string on curved D-branes
satisfies very complicated Neumann boundary conditions modified by
fermion bilinears. Thus it is in practice difficult to find the
open string mode expansion for the curved D-branes and to
generalize the worldsheet formulation for the supersymmetry
analysis in Sec. 5 to the present case. Nevertheless it is
possible to identify unbroken supersymmetries of the
curved D-branes described by Eqs. \eq{d7-curved-omega}-\eq{d3-curved-omega}
by applying the results given in \ct{mama,hikida,ggsns}.

In order to discuss the supersymmetry, it is convenient to
introduce a Fock space notation. The vacuum is defined to be
the spinor annihilated by all $b_i$'s and denoted by $|\downarrow \rangle
= (-,-,-,-)$. And the top spinor annihilated by all
$b_i^\dagger$'s is denoted by $|\uparrow \rangle = (+,+,+,+)$.
The other fourteen spinors are defined as
$|i\downarrow \rangle = b_i^\dagger |\downarrow \rangle =
(+,-,-,-)_{i=1}, \; |i\uparrow \rangle = b_i|\uparrow \rangle
= (-,+,+,+)_{i=1}$, and $|ij \rangle = b_i^\dagger b_j^\dagger
|\downarrow \rangle = (+,+,-,-)_{i,j=1,2}$ or $ b_i b_j|\uparrow \rangle
= (-,-,+,+)_{i,j=1,2}$. The IIB chiral spinors $\epsilon_\pm$
in light-cone are 8-component complex spinors. $\epsilon_+$
is related to the supersymmetries that are non-linearly realized
on the worldsheet, i.e., the kinematical supersymmetries and
$\epsilon_-$ is to those that are linearly realized
on the worldsheet, namely, the dynamical supersymmetries.\footnote{Note that
we are using the different convention from that in \ct{mama,hikida,ggsns}
according to our assignment of $SO(8)$ chirality for the
kinematical and the dynamical supersymmetries. To be with this
convention, we take the vacuum $|\downarrow \rangle$ as a negative
chirality spinor, viz., $\gamma |\downarrow \rangle = - |\downarrow \rangle$.}
It was shown in \ct{mama} that the supersymmetric
solutions $\epsilon_\pm$ for the maximally supersymmetric
background \eq{pp-metric} can be parameterized in this Fock space as
\bea \la{epsilon+}
&& \epsilon_+ = \kappa_1 + i \kappa_2 =  \beta^i \bar{Z}^i
|i \downarrow \rangle  + \delta^i Z^i| i \uparrow \rangle, \\
\la{epsilon-}
&& \epsilon_- = \epsilon_1 + i \epsilon_2 = \alpha |\downarrow
\rangle + \eta^{ij} |ij \rangle  + \zeta |\uparrow \rangle,
\eea
where $\alpha, \eta^{ij}, \zeta, \beta^i$ and $\delta^i$ are
complex constants.

\begin{table}[tbp]
\begin{center}
\begin{tabular}{|c|c|c|c|c|} \hline
D-brane type &      $W$         & $\Omega$             & $q^+$ & $q^-$\\
\hline
$D7$      &  \eq{d7-curved-brane}   &  \eq{d7-curved-omega}   &   0
& $|\downarrow \rangle, |\uparrow \rangle$     \\
\hline
        & \eq{d5-curved-brane1} & \eq{d5-curved-omega1}
        & $ |4 \uparrow \rangle,  |4 \downarrow \rangle $
        & $ |\downarrow \rangle, |\uparrow \rangle$  \\
\cline{2-5}
$D5$    & \eq{d5-curved-brane2} & \eq{d5-curved-omega2}
        & 0  & $|\downarrow \rangle, |12 \rangle, |34 \rangle, |\uparrow \rangle$ \\
\cline{2-5}
        & \eq{d5-curved-brane3} & \eq{d5-curved-omega3}
        &  $|4 \uparrow \rangle, |3 \uparrow \rangle, |3 \downarrow \rangle,
        |4 \downarrow \rangle $
        & $|\downarrow \rangle, |12 \rangle, |34 \rangle, |\uparrow \rangle$ \\
\hline
$D3$     & \eq{d3-curved-brane} & \eq{d3-curved-omega}
        & $ |4 \uparrow \rangle, |3 \uparrow \rangle, |3 \downarrow \rangle,
        |4 \downarrow \rangle $
        & $|\downarrow \rangle, |12 \rangle, |34 \rangle, |\uparrow \rangle $  \\
\hline
\end{tabular}
\end{center}
\caption{Supersymmetric curved D-branes. We indicated the equation number for
the superpotential $W$ and the gluing matrix $\Omega$.}
\label{tablefour}
\end{table}

For an open string on a curved D-brane described by a gluing matrix $\Omega$,
we have the boundary conditions \ct{ggsns}
\bea \la{bc-spinors-k}
&& \kappa_1 = \Omega \kappa_2, \\
 \la{bc-spinors-e}
&& \epsilon_1 = - \Omega \epsilon_2.
\eea
In terms of the Fock basis \eq{epsilon+} and \eq{epsilon-}, the
boundary conditions \eq{bc-spinors-k} and \eq{bc-spinors-e},
respectively, can be expressed as
\bea \la{bc-fock-k}
&& \Omega |i \downarrow \rangle = a_i |i \downarrow \rangle,
\quad  \Omega |i \uparrow \rangle = b_i |i \uparrow \rangle, \\
\la{bc-fock-e}
&& \Omega |\downarrow \rangle = \Omega |ij \rangle = \Omega |\uparrow \rangle
= \mbox{constant},
\eea
where $a_i$ and $b_i$ are complex constants and no summation for
$i$ is assumed in Eq. \eq{bc-fock-k}.
We can explicitly solve Eqs. \eq{bc-fock-k}-\eq{bc-fock-e} for
each $\Omega$ in \eq{d7-curved-omega}-\eq{d3-curved-omega}
which allows us to identify the unbroken supersymmetries. We
summarized the possible solutions corresponding to kinematical and
dynamical supersymmetries in Table 4. The spinors
$ |4 \uparrow \rangle,  |4 \downarrow \rangle $
for Eq. \eq{d5-curved-omega1} in Table 4 satisfy
$\Pi \Omega \Pi\Omega = 1$, which thus give rise to the $D_+$-brane kinematical
supersymmetry while the other two do $\Pi \Omega \Pi\Omega = -1$
giving rise to the $D_-$-brane kinematical supersymmetry. The
kinematical supersymmetry for Eq. \eq{d5-curved-omega1} is not
inconsistent with \ct{ggsns} since they concerned only
the $D_-$-brane kinematical supersymmetry satisfying
$\Pi \Omega \Pi\Omega = -1$.

There may be curved D-branes preserving only kinematical
supersymmetries on which $W$ is not constant and thus dynamical
supersymmetries are completely broken. We will not explore this
kind of curved D-branes.

\section{Discussion}

The aim of this paper was to give a systematic classification of
D-branes in the type IIB plane wave background. Of course, our
work does not mean to complete this goal even in the maximally
supersymmetric plane wave background \eq{pp-metric}.
We only considered a free string theory in the plane wave
background and static D-branes. Moreover we used the light-cone
open string theory where only longitudinal D-branes are visible.
String interactions may break certain symmetries. Furthermore
there can exist less supersymmetric solutions than those found in
this paper, e.g., by considering D-branes intersecting
at general angles \ct{bdl,ohta} or by considering more general
plane wave backgrounds \ct{mama}.

Let us discuss some related issues mentioned above. In this paper
we considered only static D-branes. However one can generate new
D-branes, symmetry related D-branes which are in
general time-dependent branes \ct{skenderis1}, from static
D-branes using the symmetries being in the action \eq{gs-action}
and the target spacetime \eq{pp-metric} but broken by D-branes,
e.g., the translation and the boost generators along the
transverse directions, $P^{r^\prime}$ and $J^{+r^\prime}$.
A rotating D-brane and a giant graviton in Penrose limit can be
described by these symmetry related boundary conditions which
preserve the same amount of supersymmetry \ct{skenderis1}.

In this paper we studied parallel and orthogonally intersecting
D-branes only. It will be straightforward to extend our analysis
to D-branes intersecting at general angles \ct{cha}. Since the rotational
symmetry is reduced to $\so \times \sop$, there are only two kinds
of supersymmetric intersection at general angles, resulting in
less supersymmetric D-brane configurations. One is generated
by $SU(2) \subset \so$ or $\sop$ and the other is generated by
$SU(2) \times SU(2) \subset \so \times \sop$. The former case
preserves the supersymmetry by half after rotation while the
latter does by quarter like in the flat spacetime.

The dynamical supersymmetry of an OD-brane also appears to be
broken when the brane is located away from the origin of
transverse space and the breaking terms depend only on the
boundary value of some Dirichlet coordinates like as $D_-$-branes.
It can be shown \ct{skenderis1}, however, that the broken dynamical
supersymmetries can be restored by modifying transformation rules
using a worldsheet symmetry realized in the action \eq{gs-action}.
However it was argued in \ct{freedman} that these extra symmetries
in the case of D-branes are not respected by string interactions.
The same thing happens for the kinematical supersymmetry preserved
by $D_+$-branes. Thus this result implies that the dynamical
supersymmetry of a $D_-$-brane or a related OD-brane located away
from the origin and the kinematical supersymmetry of a $D_+$-brane
or a related OD-brane are broken down by turning on the string
interaction.

The isometry in the plane wave background \eq{pp-metric} is indeed
$\so \times \sop \times {\bf Z}_2 $ where the first $\so$ is a remnant of
the $SO(4,2)$ isometry group of $AdS_5$ and the second $\sop$ is a
remnant of the $SO(6)$ isometry group of $S^5$. The peculiar
${\bf Z}_2 $ symmetry exchanges these two $SO(4)$'s as defined by \eq{z2}.
As we speculated in footnote \ref{z2-symmetry},
the $\bfz_2$ symmetry seems to play an important role in the existence
of OD-branes. This discrete symmetry survives only in the strict
plane wave limit. This symmetry is broken if we perturb slightly
away from the limit to $AdS_5 \times S^5$. In the pp-wave/SYM duality, the
rotation group $\so \times \sop$ in the string theory is mapped to the
product of the Lorentz (Euclidean) symmetry and the R-symmetry,
$\so_{L} \times \so_R$, in the field theory. Thus, on
the field theory side, the $\bfz_2$ symmetry interchanges the
action of $\so_{L}$ with $\so_R$. A symmetry between
spacetime and the internal (R-)space is quite novel.
Since the $\bfz_2$ symmetry is inherited from closed strings,
it is natural to expect that the $\bfz_2$ symmetry is respected by string
interactions unlike that in \ct{freedman}.
If this is true, the supersymmetry of OD-branes descending from
the closed string will be preserved even after introducing string
interactions.

It should be interesting to generalize the classification of
D-branes as done in this paper to other
backgrounds, e.g., the Penrose limit of $AdS_5 \times S^5/\bfz_N$,
the type IIA plane wave and the G\"odel universe.
We hope to address these problems in the near future.

\section*{Acknowledgments}

We are grateful to Yasuaki Hikida for useful discussions and
explaining us their work \ct{hikida}. We are supported by the grant from the
Interdisciplinary Research Program of the KOSEF (No.
R01-1999-00018) and a special grant of Sogang University. BHL is
supported by the Korean Research Foundation Grant KRF
2003-015-C00111, 2003-070-C00011, and D00027. HSY was supported by
the Brain Korea 21 Project in 2003.

\newpage

\appendix

\section{D-brane Field Equations}

Here we will show, using the general D5-brane field equations
derived in \ct{sken-tayl}, that the modified Neumann boundary conditions in
Eqs. \eq{+bc-od5} and \eq{-bc-od5} can
be realized by appropriately turning on a worldvolume flux.

In our case under consideration, $\Phi = B_{ij}= C_{n}^{RR}=0$ except
as $C_4^{RR}$ giving rise to the
background $F_5 = d C^{RR}_4$ in Eq. \eq{pp-metric}.\footnote{In this Appendix
we will use the notation adopted in \ct{sken-tayl} and set $2 \pi \a^\prime =1$.}
The relevant worldvolume Chern-Simons coupling is then of the form
\begin{equation}\label{Chern-Simons}
S_{WZ} = \tau_5 \int_{\Sigma_6} C^{RR}_4 \wedge \widehat{F}_2
\end{equation}
where $\widehat{F}_2$ is a gauge flux in the $D5$-brane worldvolume $\Sigma_6$.
The equations of motion for a $D5$-brane read as (see Eq. (3.2) in
\ct{sken-tayl}):
\bea \la{d5eom-A}
&& \p_i (\sqrt{-M} \theta^{ii_1}) = \frac{1}{5!}
\epsilon^{i_1i_2i_3i_4i_5i_6}F_{i_2i_3i_4i_5i_6}, \\
\la{d5eom-X}
&& \p_i (\sqrt{-M} G^{ij} \p_j X^n g_{mn}) = 0,
\eea
where $M_{ij} = g_{ij} + \widehat{F}_{ij}$ and
\bea \la{symm-M}
&& G^{ij}=\biggl( \frac{1}{g+\widehat{F}}g\frac{1}{g-\widehat{F}}
\Biggr)^{ij}, \\
\la{antisymm-M}
&& \theta^{ij}=\biggl( \frac{1}{g+\widehat{F}} \widehat{F} \frac{1}{g-\widehat{F}}
\Biggr)^{ij}.
\eea

Let us first check the equations of motion \eq{d5eom-A} and \eq{d5eom-X}
for the familiar example, $(+,-,4,0)\; D5$-brane, which also serves to fix
the relation between the worldvolume flux $\widehat{F}_{2}$ and
the field strength $F_{2}$ entering in the worldsheet boundary coupling
such as \eq{+bi} and \eq{-bi}. The world volume coordinates are
given by $\xi^i=(X^+, X^-, X^1, X^2, X^3, X^4)$. Since
$\sqrt{-M}=1$ and $G^{++}=0$ in the metric \eq{pp-metric}, the
left-hand side of Eq. \eq{d5eom-X} identically vanishes and thus
Eq. \eq{d5eom-X} gives no constraint. However, Eq. \eq{d5eom-A}
gives some constraint for flat $(+,-,4,0)$ embedding since
\be \la{40-embed}
\p_r \widehat{F}^{-r} = 2 \mu, \qquad (r = 1, \cdots,4)
\ee
where we used the fact that $\theta^{-r} = \widehat{F}^{-r}$. Now we
identify
\be \la{factor2}
2 \widehat{F}^{-r} = F^{-r}
\ee
from which we get\footnote{Indeed the factor 2
difference was originated from our different normalization between
worldsheet light-cone coordinates defined by
$\sigma^{\pm}= \half(\tau \pm \sigma)$ and worldvolume light-cone coordinates
given by $\xi^{\pm}= X^\pm = \frac{1}{\sqrt{2}}(X^0 \pm X^9)$.}
\be \la{40-F}
F^{-r} = \mu X^r
\ee
which is exactly the Born-Infeld flux necessary to preserve the
dynamical supersymmetry of $(+,-,4,0)$-brane \ct{kim}.

Let us now check the equations of motion \eq{d5eom-A} and \eq{d5eom-X}
for an $OD5$-brane, to be specific,
described by $\Omega = \frac{1}{2}(\gamma^{1}\!-\!\gamma^{6})
(\gamma^{2}\! - \!\gamma^{5})\gamma^{34}$ whose world volume coordinates
$\xi^i=(X^+, X^-, X^{\hat{1}}, X^{\hat{2}}, X^{\dot{3}}, X^{\dot{4}})$
are given below Eq. \eq{dia-coordinate}. Only Eq. \eq{d5eom-A}
gives a constraint for flat embedding $\xi^i$ with flux:
\be \la{od5-embed}
\p_r \widehat{F}^{-r} = \mu, \qquad (r = \hat{1}, \hat{2}
\; \mbox{or} \; \dot{3}, \dot{4}).
\ee
Note that the right-hand side of Eq. \eq{od5-embed} is reduced by
half compared to the $(+,-,4,0)$-brane since the worldvolume of
the $OD5$-brane, $\widehat{\Sigma}_6$, is slanted against the
background RR 5-form $F_5 = d C^{RR}_4$. Using the relation
\eq{factor2}, we get
\bea \label{od5+flux}
&& F^{-\rd}= \mu X^{\rd}, \qquad \rd = \dot{3}, \dot{4}, \\
\la{od5-flux}
&& F^{-\rh}= \mu X^{\rh}, \qquad \rh = \hat{1}, \hat{2},
\eea
which are the correct worldvolume fluxes necessary to preserve the
dynamical supersymmetry of the $OD5$-brane as was shown
in Eqs. \eq{+bi} and \eq{-bi}.

Similarly, we can apply the same method to an $OD_+5$-brane with
flux. If the $OD_+5$-brane could preserve the dynamical supersymmetry,
the Neumann boundary condition should be modified as
$\p_\sigma X^r - \mu X^r =0, \; \forall r = (\hat{1}, \hat{2}, \hat{3}, \hat{4})$.
However it is not possible since the equation \eq{d5eom-A}
for the $OD_+5$-brane embedding with flux now requires
\begin{equation}\label{od+5-embed}
\p_r F^{-r} = \mu \;\; \Leftrightarrow  \;\; F^{-r}= \frac{\mu}{4}
X^r.
\end{equation}

\newpage

\bibliographystyle{JHEP-like}

\end{document}